\documentclass{elsarticle}

\usepackage{fullpage}

\usepackage{amssymb}

\usepackage{boxedminipage}

\usepackage{listings}

\usepackage{ifpdf}

\usepackage{graphicx}
\usepackage{paralist}
\usepackage{placeins}
\usepackage{subfigure}
\usepackage{url}

\begin{document}
	
\begin{frontmatter}

\title{Tester Interactivity makes a Difference in Search-Based Software Testing: A Controlled Experiment}

\author[1]{Bogdan Marculescu}
\author[1]{Simon Poulding}
\author[1]{Robert Feldt}
\author[1]{Kai Petersen}
\author[2,1]{Richard Torkar}

\address[1]{Blekinge Institute of Technology, Karlskrona, Sweden}
\address[2]{Chalmers and the University of Gothenburg, Gothenburg, Sweden}

\begin{abstract}

	\textbf{Context:} Search-based software testing promises to provide users with the ability to generate high-quality test cases, and hence increase product quality, with a minimal increase in the time and effort required.

	One result that emerged out of a previous study to investigate the application of search-based software testing (SBST) in an industrial setting was the development of the Interactive Search-Based Software Testing (ISBST) system. ISBST allows users to interact with the underlying SBST system, guiding the search and assessing the results. An industrial evaluation indicated that the ISBST system could find test cases that are not created by testers employing manual techniques. The validity of the evaluation was threatened, however, by the low number of participants. 
	
	\textbf{Objective:} This paper presents a follow-up study, to provide a more rigorous evaluation of the ISBST system. 
	
	\textbf{Method:} To assess the ISBST system a two-way crossover controlled experiment was conducted with 58 students taking a Verification and Validation course. The NASA Task Load Index (NASA-TLX) is used to assess the workload experienced by the participants in the experiment.
	
	\textbf{Results: }The experimental results validated the hypothesis that the ISBST system generates test cases that are not found by the same participants employing manual testing techniques. A follow-up laboratory experiment also investigates the importance of interaction in obtaining the results.
	
	In addition to this main result, the subjective workload was assessed for each participant by means of the NASA-TLX tool. The evaluation showed that, while the ISBST system required more effort from the participants, they achieved the same performance.
		
	\textbf{Conclusions:} The paper provides evidence that the ISBST system develops test cases that are not found by manual techniques, and that interaction plays an important role in achieving that result.

\end{abstract}


\end{frontmatter}

\section{Introduction} 
\label{sec:introduction}

Software testing plays a crucial role in increasing the quality of software systems, as well as the perceived quality of and confidence in such systems. One software testing technique is the application of metaheuristic optimization algorithms to generate test data, known as Search-Based Software Testing (SBST)~\cite{McMinn2011,AfzalTF09}.

In a previous study~\cite{marculescu2012concept}, we have proposed a system that would allow successful application of SBST in an industrial context. This system, called the Interactive Search-Based Software Testing (ISBST) tool, incorporated the domain knowledge existing in the company into the search process. This was achieved by allowing human testers to interact with the system and guide the evolution of the search-based solutions. The interaction was inspired by work in Interactive Evolutionary Computation~\cite{Feldt99, Takagi2001, Feldt02, Simons2010, bush2011hyperinteractive}, and was designed to allow the testers to make their contribution, without having to deal with the complexity of the underlying SBST system.

Previous work~\cite{marculescu2014initial} focused on successfully applying ISBST in an industrial context, and  determining what were the important factors that enabled successful application. One of the findings of that study was that the ISBST tool developed test cases that were quite different from those obtained by means of manual techniques. However, the evaluation was conducted with a low number of participants and in a context specific to our industrial partner, thus making it difficult to draw conclusions about the ISBST.

This study validates those findings, by conducting a large, controlled experiment, comparing the test cases developed using the ISBST system with those developed using a manual black-box technique. The experiment was conducted with $58$ software engineering students, participants in a software Verification and Validation course at the master level. 

The experiment provides evidence that the automated system, represented by the ISBST tool, develops different test cases from the manual method. A follow-up computer-based experiment also provides evidence for the role of interaction in obtaining the results. By isolating the interaction strategy and comparing against the same search-based system without the benefit of interaction, we were able to provide evidence that interaction plays a significant role in the results obtained by the ISBST tool.

The contributions of this paper are as follows:
\begin{itemize}	
	\item Comparing test cases developed by the ISBST system and those developed by manual exploratory testing, to identify differences and similarities between them, and to determine whether or not they investigate the same type of SUT behavior. 
	\item Assessing the effect of the interaction component of the ISBST system on the outcome of the search.
	\item Widening the application of the ISBST system to a completely new type of System under Test (SUT), part of a different domain.
	\item Evaluating the ISBST system on a wider set of participants, in a controlled environment.
\end{itemize}

Section~\ref{sec:related_work} describes existing work on evolutionary approaches and search-based software testing and discusses the context of the current approach, as well as providing a description of the ISBST system itself. In Section~\ref{sec:experimental_design} we describe the design of the current experiment and the tools used during the experimental process. Sections~\ref{sec:results} and~\ref{sec:discussion} present the results from the experiment and discuss their significance, respectively. The threats to the validity of the study are discussed in Section~\ref{sec:threats_to_validity}, and Section~\ref{sec:conclusions} concludes the paper. 


\section{Context} 
\label{sec:related_work}


This experiment is inspired by results from a study conducted with our industrial partner, to investigate the possibility of using interactive search-based software testing (ISBST) to improve the testing process. Our industrial partner develops embedded software for industrial applications. The ISBST tool was developed and evaluated in that context, on a small number of company engineers. Therefore, this study will evaluate the ISBST tool outside of that specific context and with a larger number of participants. 

We define a ``domain specialist'' as a person that develops and tests software for their specific domain as part of their activities, but that is not a software engineer. To assist domain specialists, tools are specifically designed to use the terminology, symbols, and concepts specific to the domain, rather than those specific to software development and testing. Thus, they focus on domain experience and expertise rather than knowledge specific to software testing.

In previous work~\cite{marculescu2012concept} we proposed a tool, called the Interactive Search-Based Software Testing (ISBST) tool, that would use search-based techniques to help in the testing process. It is difficult to develop \emph{a priori} a fitness function that would be useful for a general SUT. As a result, the ISBST tool was designed to use a Dynamically Adapted Fitness Function (DAFF). In this concept, the fitness function is composed of a set of dimensions relevant to system quality to assess each candidate solution. By changing the relative importance of these attributes, the domain specialist can change the fitness function and indirectly guide the search. In our previous study, the relevant dimensions were identified and validated in collaboration with our industrial partner. 

Further work~\cite{marculescu2014initial} resulted in a practical implementation of the ISBST tool. The tool, and the concept of a Dynamically Adapted Fitness Function, were validated in a small case study conducted in an industrial setting. One of the results of that study was that the test cases that were developed by using the ISBST tool were useful and unexpected. The domain specialists using the tool stated that they would not have considered investigating that type of behavior, but that the behavior itself was a good addition to the test suite. 

The results of the exploratory study mentioned above indicated that using the ISBST tool would enable domain specialists to guide the search towards a more diverse set of behaviors than they could develop by using manual techniques. The more diverse set of behaviors would then be assessed by the domain specialists, who would refine relevant test cases and add them to the test suites. 

Henceforth, we define the ``behavior'' as the set of measured outputs, or any function of those outputs, corresponding to a given set of inputs of the system under test (SUT). Thus, the ``observed behavior space'', or just ``behavior space'', is the total set of possible behaviors for a given SUT\@. Note that the behavior space deals only with characteristics of the SUT that are measured or evaluated, and is not a complete description of the SUT\@. 

Henceforth, the behaviors that are measured and form the behavior space will be called ``behavior attributes''. The ISBST system may try to optimize, i.e.\@ minimize or maximize, the found values for a given behavior attribute in a direction. In this case, a ``search objective'' is defined as the combination of behavior attribute and direction. 

Additional behavior attributes may be identified and added, if they are considered relevant, and this would result in changes to the the behavior space of the SUT\@. This further complicates attempts to explore the behavior space. For this paper, we define a ``test case'' to consist of a set of inputs and the corresponding SUT behavior.  

For most of the SUTs developed by our industrial partner, there is a complex relationship between the inputs and the behaviors. Moreover, the SUTs themselves are meant to be part of a more complex product, with their behaviors becoming the inputs for other systems. This means that failures, especially failures that cannot be foreseen, propagate through the system.

In a testing context, this means that it is important to know what regions of the behavior space can be reached. It also means that testers are not able to directly explore the behavior space and determine the combination of inputs that would result in interesting behaviors. Thus, exploring the behavior space must be done indirectly, by exploring the input space. 


\subsection{Related Work} 
\label{sub:related_work}

Search-based software testing (SBST) is the application of metaheuristic optimization methods to the problem of software testing. SBST is part of the larger scope of search-based software engineering, a term coined by Harman and Jones~\cite{Harman2001}. SBST has been successfully applied on a wide range of software testing problems. McMinn~\cite{McMinn2011} describes the use of SBST for temporal, structural, and functional testing, while Afzal et al.~\cite{AfzalTF09} focus their review on the use of SBST on non-functional testing.

Search-based techniques, both in the wider area of software engineering and, more specifically in the field of testing, rely on having an automated means of assessing the quality, or ``fitness'' of a candidate solution. 

However, the definition and understanding of what fitness is, and what candidates are preferable, can change during the search. This can be the result of an evolving understanding of the problem, i.e.\ previously unknown information becomes available, or through clarifying misunderstandings or omissions, e.g.\ implicit domain knowledge not mentioned previously is now explicitly included in the fitness evaluation. As a result, designing a relevant fitness function \emph{a priori}, i.e.\ at the beginning of the process, has proven to be a challenge. 

One way of addressing this issue was to engage human users in the search, to add their knowledge and intuition to the search. The user can interact with a search-based system at several levels of abstraction. At the highest level, the user sets the target that the search should reach, and allows the automated system the freedom to find solutions. At a medium level of abstraction, the human can change the way the fitness of candidates is being evaluated. The lowest level of abstraction puts the user in a position to directly influence how the search is performed. e.g.\ how new candidate solutions are developed.

An example of the highest level of abstraction would be an automated system that develops tests as the human user writes code or defines specifications~\cite{Feldt02}. The system would be influenced by the user indirectly, by having to adjust to the constantly changing goal. Indirect interaction could also be used to explore alternative designs, to understand design constraints or to assess alternative design decisions~\cite{Feldt99, parmee2000multiobjective}.

At the medium level, a user can more directly guide the search by replacing the fitness function. Takagi proposed Interactive Evolutionary Computation (IEC), which he describes as an Evolutionary Computation (EC) ``that optimizes systems based on subjective human evaluation''~\cite{Takagi2001}. This would allow the human user to guide the search according to their ``preference, intuition, emotion and psychological aspects''~\cite{Takagi2001}. IEC could then see a wider spectrum of applications, including arts and animation. 

Alternatively, a system may require the human user to only replace the fitness functions at certain times, e.g.\ to serve as a tie-breaker, when the existing fitness functions cannot rank certain candidates~\cite{Tonella:2010:UIG:1915081.1916189}. 

The fitness function itself can be subject to change, including user preference as a factor in computing fitness~\cite{4804706, 5585740}, having elegance as a key factor in software design~\cite{simons2012elegant}, or readjusting the fitness function to ensure that user preferred candidates receive a higher fitness score~\cite{liapis2012limitations}.

At the lowest level, interaction can be very detailed. Bush and Sayama~\cite{bush2011hyperinteractive} require the human to be ``the main driver of the search process'' by selecting the individuals and the evolutionary operators to be applied. 

Replacing the fitness function with a human user, however, makes the fitness evaluation subjective and dependent on the individual user. This is not a problem for applications where subjective impressions are key, such as art, but might raise concerns when applied to engineering problems. 

A more serious problem is that the number of evaluations that a human can perform is limited, as boredom and fatigue will set in. This is even more of an issue at the lowest level of abstraction, where the human user is involved in evolving each candidate. Fatigue has already been identified as a major concern, and efforts to alleviate the problem have been proposed~\cite{Kamalian2006}. Alternatives have, therefore, been proposed that make it easier for the human user to interact, by selecting candidate solutions they favor and dismissing those they do not~\cite{walsh2010terrain}; or focusing on the search objectives or the fitness values more than on the candidate solutions themselves~\cite{hayashida2000visualized, Bavota:2012:PDI:2415185.2415195}.

Existing work on interaction in evolutionary computation seems to be focused on areas other than software testing. Nevertheless, the interaction techniques being described are applicable on any search-based system, as long as elements of it are subject to human evaluation. Moreover, human preference can help guide the process where the objectives of the search are unknown or unknowable. This is evident in applications such as aesthetics and software design, but ambiguity exists in other areas as well.


\subsection{Interactive Search-Based Software Testing (ISBST)} 
\label{sub:interactive_search_based_software_testing_isbst_}

The ISBST tool is designed to make it easy for a domain specialist to support the search process with their knowledge and experience, without requiring familiarity with the particular implementation of the underlying algorithm. In terms of the levels being described above, the ISBST system exists at the higher level. Users of the system will interact with the system to develop the fitness function and provide an evaluation of some of the resulting test cases, but not replace the fitness function or evaluate each individual test case. This approach is aimed at allowing the user to control how the search proceeds for the entire population, rather than focusing in on individual candidate solutions. 

The ISBST system generates, based on guidance from the domain specialist, a population of candidate solutions or ``candidates''. An overview of this population of candidates is provided to the domain specialist. The domain specialist can select from the population candidates that are of interest, obtain more information about them, and export them for use. The domain specialist can also change the goals of the search, in order to guide the search towards interesting system behaviors. From the current population, and with the goals set by the domain specialist, the search resumes.

\begin{figure}
	\centering
		\includegraphics[scale=0.6]{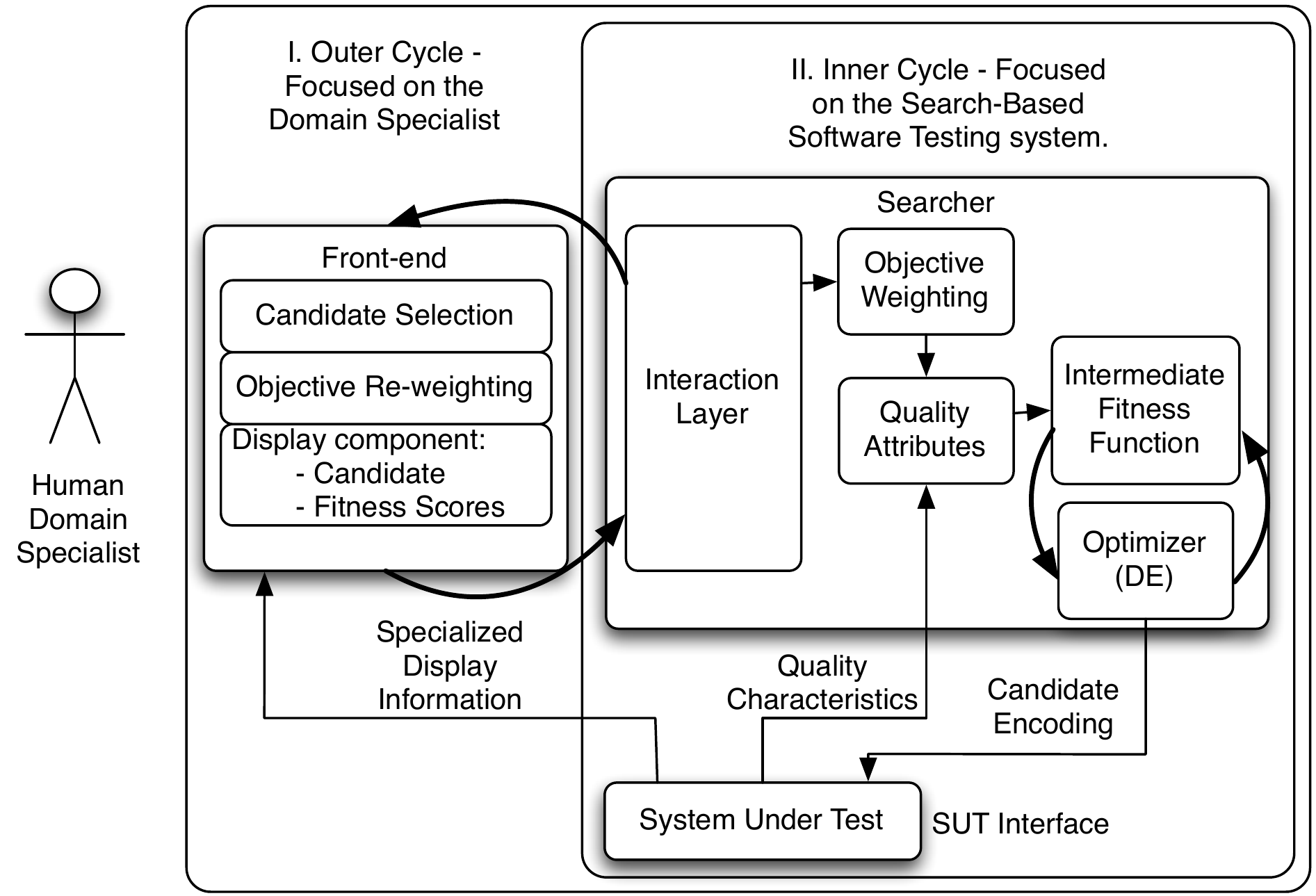}
	\caption{Overview of the ISBST System}
	\label{fig:ISBST}
\end{figure}

The system is composed of two nested components, the \textit{inner cycle} that contains all the components for initiating, running, and guiding the search, and the \textit{outer cycle} that handles the interaction with the user, as shown in Figure~\ref{fig:ISBST}. The \textit{outer cycle} interacts with the user periodically, displays the candidates, collects the inputs, and then resumes the search with the new input. Between two interactions, the \textit{inner cycle} is run with the selected inputs.

	\textbf{The \emph{Inner Cycle}} consists of the search algorithm itself, the means of interacting with the SUT, and the means for guiding the search. When the search algorithm generates a new test case candidate, that candidate consists only of a set of inputs. The inputs are then fed into the SUT, to obtain the corresponding behavior. Once the candidate is complete, its fitness is evaluated by means of a fitness function.

Defining a fitness function \emph{a priori} is a difficult task, as it is impossible to guess the specific details of each system and each situation. To handle search guidance, the ISBST system uses a number of behavior attributes. The behavior attributes are generally SUT specific or, at most, domain specific. In our experience so far, behavior attributes have been defined during the development process, relying on input from domain specialists, and validated with the help of domain specialists. Each behavior attribute has a weight, set by the domain specialist via the \emph{outer cycle}. The weight represents the importance of each behavior attribute with respect to the others. 

The fitness of a candidate is a weighed sum of the scores obtained by that candidate for each of the behavior attributes: 
\begin{equation}
	DAFF_{j} = \sum_{i=1}^{\mathrm{nObjectives}} { \mathrm{Weight}_{i} * \mathrm{Value}_{i,j} }
	\label{eq:iff}
\end{equation}
where $DAFF_{j}$ is the fitness value of candidate $j$, $\mathrm{Weight}_{i}$ is the current weight of the objective $i$, and $\mathrm{Value}_{i,j}$ is the fitness value of candidate $j$ measured by objective $i$. The value of $DAFF_{j}$ is the sum of the weighted fitness values for all $nObjective$ objectives. An objective $k$ can be deselected from the computation by having $\mathrm{Weight}_{k}=0$.

The relative importance of each behavior attribute is the mechanism by which the domain specialists can influence the search: values for the weight of each objective are set by the domain specialists in the \emph{outer cycle}, and then passed to the \emph{inner cycle} and used in the fitness evaluation.

	\textbf{The \emph{Outer Cycle}} enables the domain specialists to interact with the ISBST system. This interaction has two components: visualizing candidate solutions and guiding the search.

The domain specialist is shown a summary of the current population of candidates solutions and overview of the scores obtained by the candidates for each behavior attribute. Each candidate's scores for each behavior attribute, and detailed information regarding each candidate are available on demand. This information is domain and even SUT specific.

The user guides the search by setting the importance of each behavior attribute with respect to the others. This is done by assigning a weight to each behavior attribute. The weight is passed to the \emph{inner cycle} and used to compute the DAFF, and therefore guide the search. 

This guidance is achieved by allowing the domain specialist to adjust the relative importance of each of the search objectives. The objective, with their assigned weights, then form the DAFF that is used for the next set of optimization steps. After a number of optimization steps, the domain specialist is shown the latest generation of test case candidates, provided with all the information available on those candidates, and offered the possibility to adjust their weighting accordingly. This interaction with the specialist is called ``interaction event''.

The exact purpose of each test case is determined by the domain specialist, as is its fitness for that purpose. The ISBST tool generates the input data for the test cases, computes the SUT behavior corresponding to those inputs, and then evaluates the fitness of the test case according to the weighting provided by the user. As the weighting changes, so does the optimization objective of the ISBST tool.

The ISBST tool does not require the availability of an oracle to specify if the observed behavior is acceptable or not. Since the goal is to investigate areas of the input and behavior space that human domain specialists would not have otherwise considered, it can be expected that other mechanisms to assess that behavior, e.g.\ specifications, models, etc., are not available. The expectation is that domain specialists would identify those of the generated test cases that are remarkable in some way and assess that behavior themselves. 

Once a candidate, or group of candidates, has been found to be suitable, they can be exported for later review and for inclusion in test suites. The search process can then resume.


\subsection{The ISBST Tool Implementation} 
\label{sub:the_isbst_tool_implementation}

The system is implemented as a distributed system, with the \emph{inner cycle} being implemented in the Julia language\footnote{http://julialang.org/} and deployed on a remote server. The \emph{outer cycle} is implemented as a browser based application in Javascript, using the Data-Driven Document (D3) library\footnote{http://d3js.org/} to enable an informative candidate display. Communication between the two sides is done by packaging candidate objects in a JavaScript Object Notation (JSON) file\footnote{http://www.json.org/}. The system is run as a server application, and experimental participants need only connect to it via browser.

\begin{figure}
	\centering
		\includegraphics[scale=0.9]{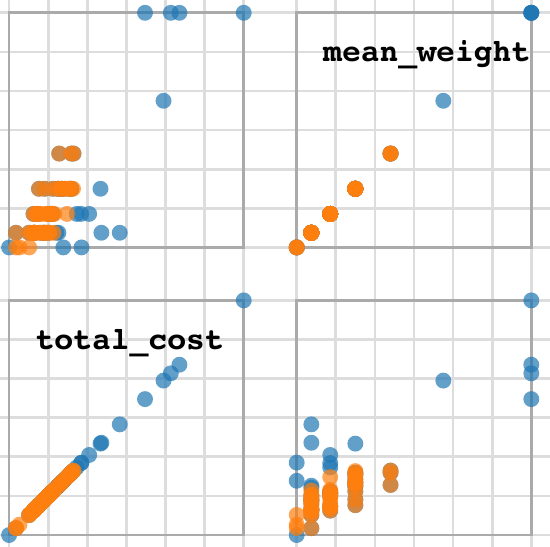}
	\caption{Example of an overview of the solutions found by the ISBST system (current generation in blue, previous generation in orange). The graph shows an example of a candidate population, plotted according to two of the relevant behavior attributes (total\_cost and mean\_weight, in this case). The matrix of scatterplots can be extended to reflect a large number of behavior attributes.}
	\label{fig:multi}
\end{figure}

\begin{figure*}
	\centering
		\includegraphics[scale=0.9]{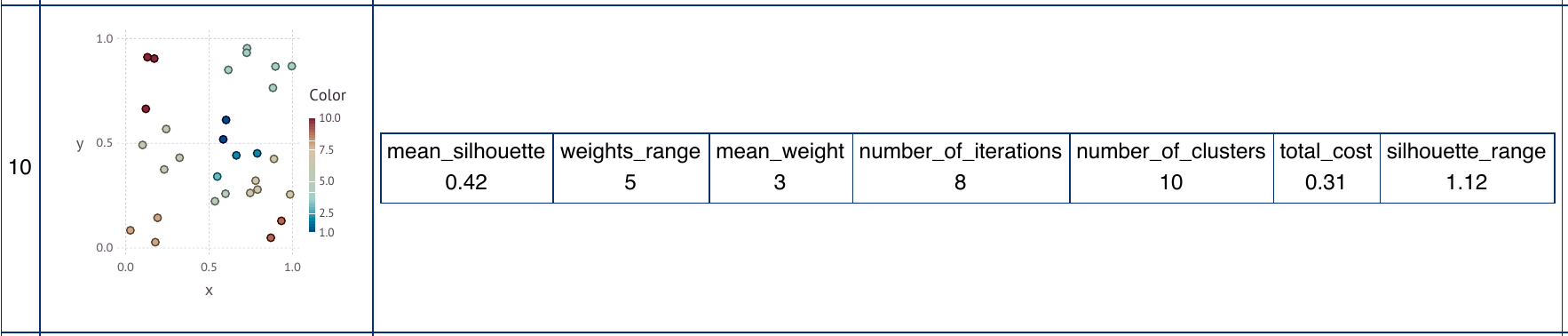}
	\caption{Detailed view of one candidate solution. The detailed view shows the score obtained by the candidate for each of the behavior attributes. This panel shows the raw values for each score, rather than the normalized values used in the calculations.}
	\label{fig:detail}
\end{figure*}

\begin{figure}
	\centering
		\includegraphics[scale=0.2]{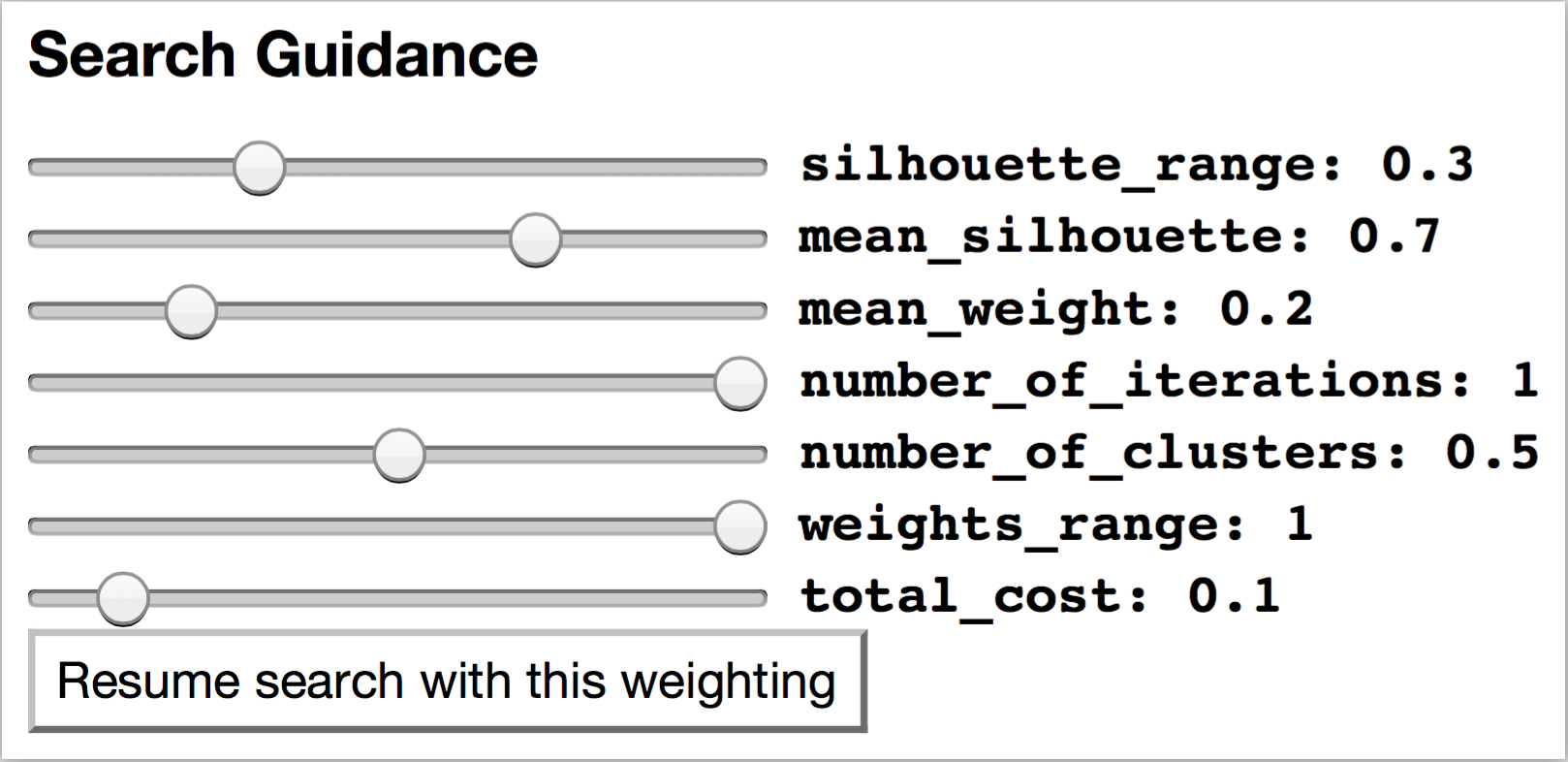}
	\caption{Detailed view of the search guidance panel. All the behavior attributes are specified here, and the domain specialist can view and alter the weight of each attribute to reflect their relative importance at any given moment.}
	\label{fig:guidance}
\end{figure}

During an interaction event, the user connects to the \emph{outer cycle} via the web page. There they can see an overview of the candidate solutions, plotted according to the scores they obtained for each quality objective, as shown in the example in Figure~\ref{fig:multi}. Detailed information for each candidate solution is also available on demand, for example in Figure~\ref{fig:detail}, to reduce the risk of overwhelming the user with information. To guide the search, the users set the relative importance (weight) of each behavior attribute by means of a set of sliders. An example of that panel of sliders can be seen in Figure~\ref{fig:guidance}.

For the current version of the system, the search algorithm used is a Differential Evolution algorithm~\cite{Storn:1997:DEN:596061.596146}. Differential Evolution is a parallel direct search method, where each potential solution is encoded as a vector of real numbers. In our case, each vector represents the input data in one test case. The initial population is chosen randomly from a uniform distribution, and covers the entire parameter space. New parameter vectors are added by mutation: adding the weighted difference between two population vectors to a third vector.

For each target vector $x_{i,G}$, where $i=1, 2, \ldots, NP$ a mutant vector is generated as follows:

\begin{equation}
	v_{i,G+1} = x_{r_1,G} + F * (x_{r_2,G} - x_{r_3,G})
	\label{eq:de}
\end{equation}

where $r_1, r_2, r_3 \in {1, 2, \ldots, NP} $, are integers, and mutually different, and different from the running index $i$. $F$ is a real and constant factor $\in (0, 2]$ which controls the amplification of the differential variation $(x_{r_2,G} - x_{r_3,G})$. If the mutant vector is an improvement over the target vector, it replaces it in the following generation~\cite{Storn:1997:DEN:596061.596146}.

The crossover rate we used is $cr=0.5$, the scale factor is $F=0.7$, and the population size is $population=100$. The mutation strategy is that proposed by Storn and Price~\cite{Storn:1997:DEN:596061.596146}: DE\slash rand\slash 1\slash bin. The strategy uses a differential evolution algorithm (DE); the vector to be mutated is randomly chosen (rand); one difference vector is used (1); the crossover scheme is binomial (bin).

To allow the single objective DE to handle multi-objective and many-objective problems, we used the Sum of Weighted Global Averages~\cite{Bentley_findingacceptable}. Individuals that have higher fitness scores in more important search objectives, i.e.\ that have a higher weight, will receive a better overall fitness score. This ensures that the search favors those individuals that the users, by means of the weights, have decided are important.

The search objectives may represent a large number of different measurements and have very different scales. The fitness value for each search objective is normalized based on the extreme values for that objective, as shown in Equation~\ref{eq:swgr}. This ensures that a search objective cannot influence the fitness function by virtue of its scale, rather than its relative importance as assessed by the user. 

The fitness ratio of candidate $i$ for one behavior attribute is shown below, in Equation~\ref{eq:swgr}.

\begin{equation}
	\mathrm{fitness\_ratio}_i = \frac{(\mathrm{value}_i - \mathrm{min}_i)}{(\mathrm{max}_i - \mathrm{min}_i)}
	\label{eq:swgr}
\end{equation}

where $\mathrm{fitness\_ratio}_i$ is the fitness ratio of candidate $i$ with respect to one behavior attribute, $value_i$ is the raw score obtained by candidate $i$ with respect to the current behavior attribute, and $min_i$ and $max_i$ are the globally best and worst values seen for the current objective in the entire search.



\section{Experimental Design} 
\label{sec:experimental_design}


The purpose of the experiment is to evaluate the ISBST system in a controlled laboratory setting. We hypothesize that the ISBST system finds test cases that are not developed by human users alone, and therefore results in different behaviors of the SUT\@. 

The experiment was designed to answer the following research questions:
\begin{itemize}
	\item \textbf{RQ1.} To assess the degree to which test cases developed by the ISBST system investigate different areas of the behavior space and, thus increase the diversity of available candidates.
	
	The main hypothesis of the experiment is that test cases developed using the ISBST system are different in terms of the SUT behavior they cover from test cases obtained using the manual black-box testing technique. 
	
	While we acknowledge that test case diversity does not necessarily imply increased fault finding ability, this remains a reasonable objective. Work by Feldt et al.\@ links test suites with higher diversity to higher structural and fault coverage~\cite{cognitively_diverse, diameter}. 
	
	\item \textbf{RQ2.} Do both the search and the interaction components of the ISBST system have a significant contribution to the observed effects? 
	
	Our hypothesis is that both components of the ISBST system have a contribution to the result and, thus, both are relevant. To answer this question, we executed a second experiment, to evaluated the impact of the search and interaction components in isolation. To achieve this goal we ran the ISBST system again, with the same settings, but without the interactive components. 
		
	\item \textbf{RQ3.} Is the ISBST system more demanding of the domain specialist than using the manual exploratory testing technique?
	We wanted to investigate in greater detail the demands that the ISBST tool places on the domain specialist. The ISBST tool adds another layer of abstraction, so we hypothesize that it will also place a greater strain on the domain specialist. 
	
\end{itemize}

The independent variable is the method being used: manual or ISBST\@. Both methods are supported by tools that make the same information available to the participant, and computed in the same way. Thus, any difference being observed, will not be due to data availability, or differences in the algorithms for behavior computation.

The dependent variables are: the set of test cases produced by each of the methods in the two experiments and the auto-assessed level of demands placed on the participants.


\subsection{Participants} 
\label{sub:participants}

The participants to the experiment were $58$ students from a software engineering master's program at the Blekinge Institute of Technology, in Sweden. All the participants were students in the Verification and Validation course, and were recruited for the experiment through the course. All the participants were volunteers. The incentive provided during the recruitment process was the opportunity to use some of the techniques taught during the course in a more practical setting. No other rewards, financial or course credits, were offered to the participants. 

The Verification and Validation course aims to teach students the importance of systematic verification and validation of software, and provides them with knowledge of available methods and techniques, complete with their capabilities and limitations. The course covers methods such as reviews, unit testing, coverage, statistical approaches, system and integration testing, reliability, and performance. During the course, students are encouraged to critically reflect and discuss topics in verification and validation, as well as to critically evaluate the strengths and weaknesses of verification and validation techniques. In addition, students gain practical experience in planning and applying test strategies and tool on open source systems and conducted automated source code inspections. 

The experiment was conducted at the end of the Verification and Validation course, so participants benefited from the knowledge obtained during that course, in particular from the practical exercises. That being said, no additional guidance was provided by the experimenter, and the way participants approached the task was deliberately left up to individual decision. 

As students in the Verification and Validation course, the participants were familiar with the manual testing technique and had been introduced to ISBST\@. Thus they were motivated to participate and had the capabilities to complete the tasks.


\subsection{System under Test} 
\label{sub:system_under_test_and_goals}

The system under test (SUT) chosen for this experiment is the Julia language\footnote{http://julialang.org/} implementation of a $k$-means clustering algorithm. The implementation is available as a Julia library\footnote{https://github.com/JuliaStats/Clustering.jl}, complete with documentation\footnote{http://clusteringjl.readthedocs.org/en/latest/}.

This system was chosen for two reasons. The first reason is the level of complexity. We define complexity both quantitatively - as the number of inputs and outputs that form each potential solution, and qualitatively - as the difficulty a participant may face in evaluating proposed solutions. 

The chosen system had to be simple enough to allow for exploratory testing within the limited time available. A more complex system than the one chosen would have a detrimental impact on the performance of the manual method and make comparison more difficult. At the same time, the system had to be complex enough to be representative of the embedded software that was the inspiration for the experiment. 

The second reason has to do with assessing the outcome of the SUT\@. For the embedded systems that served as inspiration for this study, a human expert is considered to have the most competence and experience to evaluate a proposed solution.

For the clustering SUT chosen for this experiment, we have quantitative measurements available to assess the proposed solutions as well as the human participants providing a qualitative ``sanity check''. This combination allows us to assess both methods, both in terms of the human participants' perception of solution quality and in terms of the quantitative evaluation of the different objectives. For each test case candidate, the testers can see the scores for each of the quantitative measurements and can see a graph of the clustering results. Based on that information, they decide if that is a interesting enough test case to add to the test suite or not.

In practical terms, for each test case, the set of inputs consists of $60$ points in a two-dimensional space, and the desired number of clusters. An example of the inputs can be seen in Figure~\ref{fig:manual_inputs}. The SUT is then run with these inputs, and the behavior of the SUT is the set of values obtained for each of the fitness attributes.

\begin{table*}
	\scriptsize
	\begin{tabular}{ | p{0.05\linewidth} || p{0.2\linewidth} | p{0.4\linewidth} | p{0.2\linewidth} | }
		
		\hline
		No. & Behavior Dimension & Description & Direction \\
		\hline
		\hline
		 
		1 & Number of Clusters & Number of clusters to be found. (The $k$-means algorithm required the number of clusters to be found as an input) & Minimize \\
		\hline

		2 & Number of Iterations & $k$-means clustering is done in several iterations, until the clusters are stable. & Minimize \\
		\hline
		
		3 & Mean Silhouette & The Silhouette is a quantitative way to measure how well each item lies in its own cluster. The Silhouette of each point has a value between 0 and 1, with higher values (closer to 1) indicating that the point lies well within its own cluster and there is no meaningful alternative cluster it could be assigned to. The mean is computed to provide an overview of how well the points belong to their respective clusters across the entire population. & Maximize \\ 
		\hline
		
		4 & Silhouette Range & This is the absolute distance between the lowest and the highest silhouette values found in the current candidate. A high value for this attribute means that the test case contains both well-defined and ill-defined clusters. A low value indicates that the test case contains only one of the two options. & Maximize \\ 
		\hline
		
		5 & Mean Weight & The weight of a cluster is the sum of the weights of all the points within it. For the purpose of this experiment, each point has the same weight: $weight = 1$. Larger clusters, with more widely dispersed points, will get high values for this quality objective. Small, tightly packed, clusters will get low values for this objective. & Minimize \\
		\hline
		
		6 & Weight Range & Test cases containing a combination of large and small clusters will obtain a high value with respect to this objective. & Maximize \\
		\hline

	\end{tabular}
	\caption{An overview of the Search Objectives used in this experiment.}	
	\label{tab:quality_objectives}
\end{table*}

The behavior of the SUT, for this experiment, consists of the behavior attributes shown in Table~\ref{tab:quality_objectives}. The behavior attributes are based on measurements developed to assess and validate clustering results. The aim, whether the behavior attribute is to be minimized or maximized, was arbitrarily chosen for the experiment.


\subsection{Research Instruments} 
\label{sub:manual_tool}

\textbf{The Measurement Aid.} The manual technique used is an exploratory black box testing approach, supported by a task specific tool. 

Manually writing up the points required for a test case could get tiring and frustrating, and thus impact the evaluation. To ensure that participants using both methods have access to the same type of information about the test cases being developed, and to mitigate the risk of fatigue or frustration unduly affecting the results, a manual testing helper tool was also developed. 

Participants to the experiment are expected to manually develop the inputs to the test case, i.e.\ the $60$ points to be clustered and the number of clusters. The SUT is then run with the selected inputs and the behavior of the SUT is shown. The tester decides the inputs, and then can see the resulting scores for each behavior attribute and the graph of cluster assignments. Based on that information, they can decide whether to save the test case or not.

\begin{figure}
	\centering
		\includegraphics[scale=0.9]{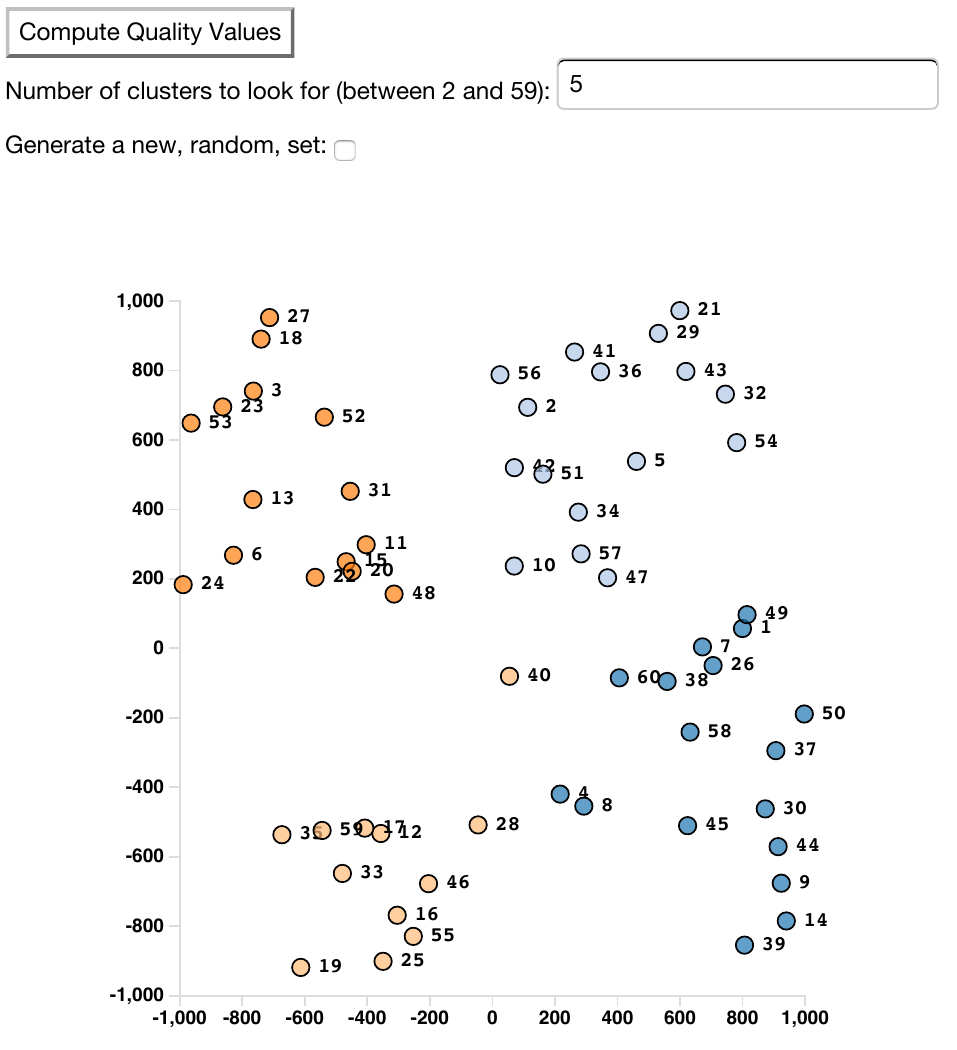}
	\caption{Defining the inputs on the manual tool.}
	\label{fig:manual_inputs}
\end{figure}

The tool allows for an easier way to draw up the test case, to view the behavior of the SUT, and to save interesting test cases, using the panel shown in Figure~\ref{fig:manual_inputs}. Participants are provided with a more convenient and intuitive way to arrange the inputs for the clustering algorithm. The Measurement Aid randomly generated the required points. Then, the participant can drag an drop any of the points in the two-dimensional input space, to match the desired set of inputs. The coordinates of the point are then updated. Once all the points are where the participant requires them to be, the inputs are sent to the SUT, the behavior is computed, and the test case is displayed to the participant. 

As mentioned previously, the behavior system consists of more than the cluster assignment, but also includes various means of assessing and describing the result. The output information presented to participants using the manual tool is the same as the information presented to those using the automated tool, and displayed in the same manner. This ensures the two methods are comparable, and that neither method has access to more information that the other. 

\textbf{NASA Task Load Index.} The NASA Task Load Index (NASA-TLX) is a subjective workload assessment tool described by Hart and Staveland~\cite{hart1988development}. The NASA-TLX is commonly used to subjectively assess the workload when working with human-machine systems~\cite{tlx-measures} in general and on assessing task difficulty in software development~\cite{tlx-difficulty} in particular.

Participants were provided with a paper and pen version of the tool, to be filled in twice by each participant: after each of the two experimental sessions. The most important dimensions for the purpose of this experiment were: mental demand, effort, temporal demand, frustration, and performance. 

The results for each dimension were analyzed separately, rather than being unified into a single demand rating. This allowed us to gain a more detailed understanding of how each method was perceived by the participants. It also allowed us to avoid uncertainties relating to the relative importance of each of the dimensions.


\subsection{Experimental Process} 
\label{sub:planned_process}

\begin{figure}
	\centering
		\includegraphics[scale=0.3]{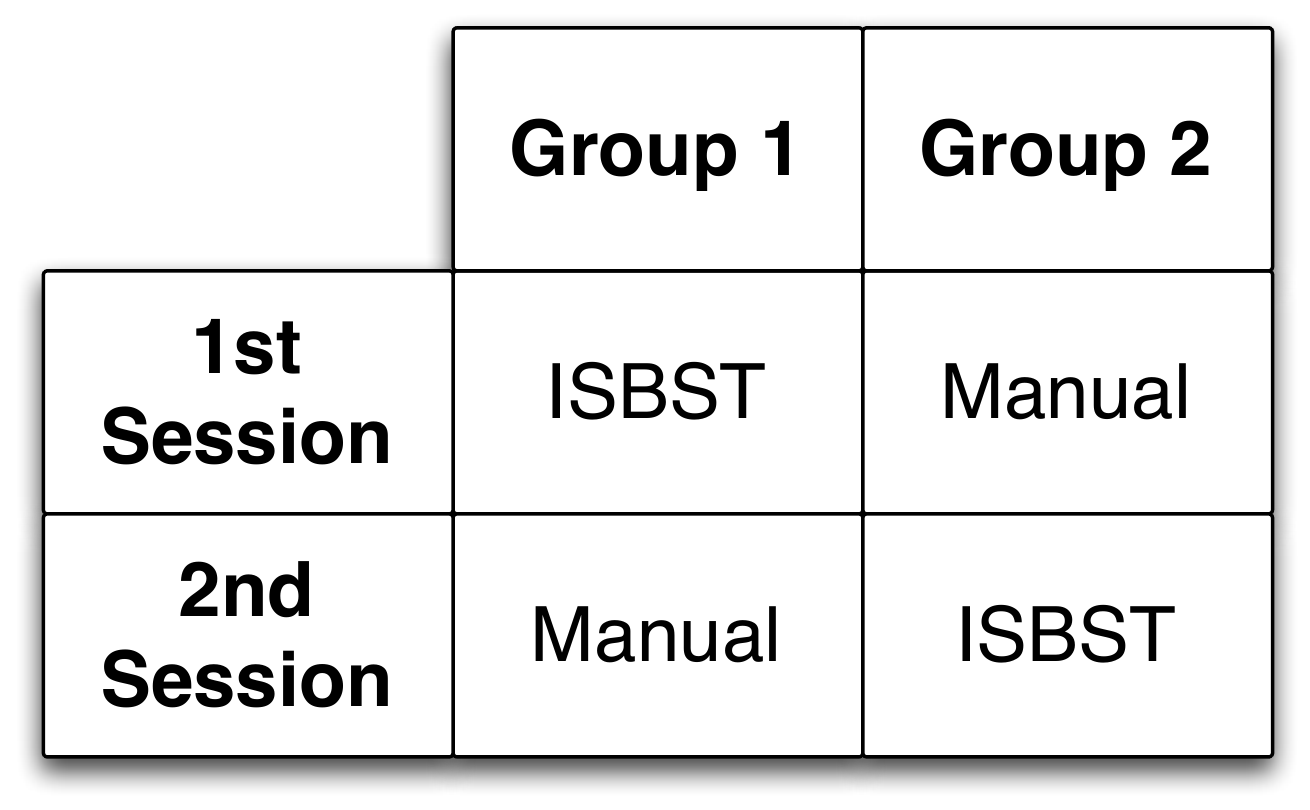}
	\caption{Overview of the Experimental Design.}
	\label{fig:crossover}
\end{figure}

The experiment is a crossover design, as shown in Figure~\ref{fig:crossover}. The experiment consists of two treatments: using the ISBST system and the Manual technique, noted in Figure~\ref{fig:crossover} as ``ISBST'' and ``Manual''. The participants are randomly split into two groups, and the experiment consists of two sessions. One group, Group 1 in Figure~\ref{fig:crossover}, will use the ISBST system for the first session and the Manual method for the second, with the second group doing the reverse. This ensures that all participants have a chance to use both techniques. 

The experiment consists of two treatments, both aimed at developing a diverse set of test cases. The first technique is the ISBST system described in Section~\ref{sub:interactive_search_based_software_testing_isbst_} in detail in our previous work~\cite{marculescu2012concept, marculescu2014initial}. 

For the second treatment, test cases are developed manually. Each participant selects the inputs for the SUT, i.e.\@ the $60$ points and the desired number of clusters. The tool described in Section~\ref{sub:manual_tool} then executes the SUT with those inputs and returns the behavior of the SUT.  

The design of the experiment was refined and validated with a pilot experiment, conducted with other researchers. The pilot was full length and allowed experimental procedures, documents, and the timing of the process to be refined. 

Due to limitations in the available laboratory space, the experiment was conducted in three separate experimental instances, all conducted within 10 days. The participants were assigned to the experimental instances randomly.

Each experimental instance began with a 20-minute presentation and demo. The presentation provided an initial description of the experiment, of the tools that would be used, and of the system under test. It also included a detailed guide describing how the participants can develop test cases using each tool and explaining the data each tool provides. The demo showed participants how to use the two tools and what they should expect from them. The presentation also emphasized the goal of the participants: to create a regression test suite that covers a wide range of interesting behaviors. The emphasis on covering many interesting behaviors is necessary, since the goal is to evaluate areas of the behavior space that each of the method covers. 

The experimental instance consisted of two 45-minute sessions. During a session, a participant would use the technique assigned to them to develop test cases. Test cases that were considered useful or interesting by the participants were then saved. 

All participants used both of the methods, with the order in which each participant used the methods being randomly assigned. Each session concluded with completing an evaluation form based on the NASA-TLX. The two experimental sessions were separated by a 15-minute break. 

In addition, each participant had a few additional documents available. First, a document providing a brief description of the behavior attributes that constitute the SUT output, an explanation of the practical consequences of high or low values for each attribute, and what the goal that the automated method was trying to reach for each attribute. A second document provided information on the method each participant used, information on how to interact with each tool, and other practical information needed to successfully complete the experiment. 

Since there was a break of a few days between experimental instances, as well as a short break between experimental sessions, we have asked participants not to discuss the details of the tasks, or the approach they used, with their colleagues. We have also made it clear to the participants that their performance on the experiment would not be used as an assessment for the course and would not impact their grades, to reduce the incentive participants had to obtain additional information from their colleagues. It was also made clear to the participants that there is no ``right'' or ``wrong'' answer that they are meant to find. 

The experimental sessions were conducted under observation, with the same researchers being involved in all sessions. This was done to ensure that all experimental sessions received the same instructions, presentation, and information. Participants has the opportunity to ask questions, and care was taken that clarifications were not leading the participants to expected behaviors.


\subsection{Data Collection and Analysis} 
\label{sub:data_collection_and_analysis}

\textbf{RQ1} is concerned with assessing the degree to which test cases developed by the ISBST system differ, in terms of the SUT behavior they explore, from test cases developed using the manual technique. The null hypothesis in this case is that there is no significant difference between the regions of the behavior space that each method explores. 

To answer \textbf{RQ1}, we collected the test cases developed by the participants during the experimental sessions and specifically marked for ``export'' or saving. This allows participants to only include test cases they regard as interesting or novel enough to consider. Since interaction with the domain expert is one of the key attributes of the ISBST system~\cite{marculescu2012concept}, so, in answering \textbf{RQ1}, we only considered test cases that participants had decided to select.

To understand the complex and extensive data that we collected, we use used several analysis techniques. First, we evaluated whether there was a statistically significant difference between the ISBST and the manually developed test cases, in terms of the regions of the SUT behavior space they investigate. To determine if the results were statistically significant we used a non-parametric test: the Mann-Whitney U-test, as suggested by Arcuri and Briand~\cite{briand-stats}. Effect sizes were calculated and interpreted using the method proposed by Vargha and Delaney~\cite{vargha2000critique}.

In addition, we wanted to look at which areas of the behavior space were explored by each method. To achieve this, we clustered the test cases based on their behavior, and analyzed the composition of the clusters. If a cluster contains test cases resulting from one method only, we can conclude that that area of the behavior space is only explored by one of the methods. 

Principal component analysis was used to isolate the behavior attributes that accounted for most of the variation. This type of analysis would highlight differences in the regions of the behavior space covered by each experimental treatment and could be used to confirm that any results are caused by the different characteristics of the methods, and to strengthen confidence in the other analysis methods. The dimensionality reduction provided by the PCA also allows visualization of the two groups of test cases so that any overlap, or lack of an overlap, can be more directly judged.

\textbf{RQ2} could be answered by a comparison of the test cases developed by the participants using ISBST system against those developed by the ISBST system without the benefit of interaction. The ISBST system without interaction assumed that all objectives has the same importance and assigned all the objectives the same relative weight.

To make sure that such a comparison would not be influenced by the any biases in the test cases participants exported, we collected the entire population of test cases developed by the ISBST system, independent of whether they were selected for export or not. The Mann-Whitney U test was used for the comparison, and effect sizes were calculated and interpreted using Vargha-Delaney.

We also collected information regarding the interaction between each participant and the ISBST tool: the number of interactions and the weights each participant used for the objectives in each interaction.

To answer \textbf{RQ2} we had to isolate the search-based system from the interactive component and assess their performance individually. A laboratory experiment was conducted to evaluate the effect of the search and the interaction separately. This laboratory experiment compared the population of test cases resulting from the realistic interactions with those obtained from a running the ISBST system in the absence of interaction. 

The realistic interaction data consists of the populations of test cases automatically collected at the end of the experiment. For each participant, the final population of $pop_{size} = 50$ test cases was recorded, as was the number of interaction events each participant used. 

For each participant, a separate search was started, and the system was run for the same amount of interaction events as the participant had used in the practical experiment, and using the same settings. The only difference was that the interaction strategy each participant used was replaced with a Null Strategy. The Null Strategy consists of keeping every objective weight to the same, non-zero, weight. This results in all the objectives have equal priority, and is equivalent to a search conducted with no interaction. We will refer to the experimental runs using the Null Strategy as ``non-interactive executions'' of the ISBST system.

This means that both the practical experiment and the non-interactive execution had the same number of optimization steps available, and thus the same number of fitness evaluations. This method of assessment, recommended by  \v{C}repin\v{s}ek et al.\@~\cite{Crepinsek:2013:EEE:2480741.2480752}, allows for a fair comparison between search algorithms.

To answer \textbf{RQ3}, as well as to get an overview of the participants to the study, information was also collected regarding their experience, skills, strategies, performance, and the level of fatigue incurred by using the methods.

Descriptive statistics, conducted on the participant data, provided an understanding of the participants' level of expertise, experience, their strategies, performance and level of demand. This approach provided insight into the degree to which the two methods are comparable in terms of the expertise and effort required.



\section{Results} 
\label{sec:results}

\subsection{Participants} 
\label{sub:rez_participants}

\begin{figure}
	\centering
		\includegraphics[scale=0.55]{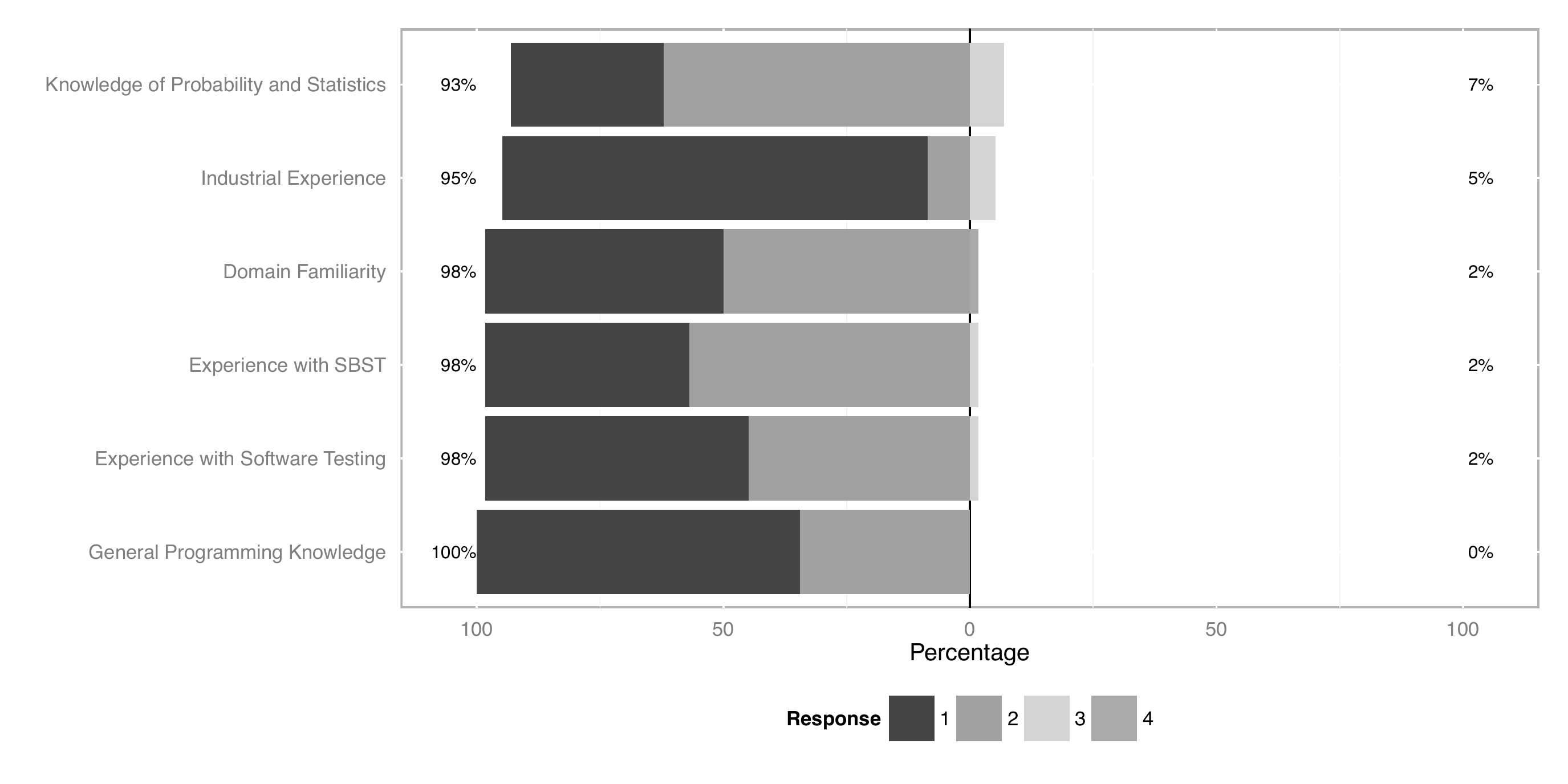}
	\caption{Results of the participant characterization test. The plot centers on the middle (between response 2 and 3). The percentages on the left are the sum for responses 1 and 2, and percentages on the right are the sum for responses 3 and 4.}
	\label{fig:participants_initial}
\end{figure}

The experiment began with a participant characterization survey, containing a number of questions aimed at assessing the participants' experience and knowledge of several key areas: general programming knowledge, industrial experience, knowledge of software testing, familiarity with SBSE, knowledge of statistics, and experience with the domain chosen for the system under test. These factors were self-assessed by the participants and the results are shown in Figure~\ref{fig:participants_initial}.

It is worth pointing out that the scales for each of the assessed dimensions were different. The results of the pre-test can be interpreted as follows:

\begin{itemize}
	\item \textbf{General Programming Experience}. This was assessed in terms of the number of programming courses the participants had taken up to the time of the experiment. The values are 1 - for one or two courses, 2 - for 3 or more courses, with higher number for practical experience in industry. Two thirds of the participants had had one or two courses (value 1 - in Figure~\ref{fig:participants_initial}), with the remaining third having had more than 3 programming courses (a value of 2 in Figure~\ref{fig:participants_initial}).
	\item \textbf{Industrial Experience in Programming}. For this dimension, answers range from 1 - no industrial experience programming, to 3 - more than one year of industrial experience. Most participants to the experiment were students with no industrial experience. 
	\item \textbf{Software Testing Experience}. This dimension assesses the experience participants had with software testing before the Verification and Validation course, with 1 representing very little to no experience in testing (even in courses), and 4 representing more than 1 year of industrial experience with testing activities. Most participants had not undertaken any explicit and systematic testing activities. We can only conclude that the Verification and Validation course was the first contact most participants had with systematic testing activities.
	\item \textbf{Experience of Search-based Software Testing (SBST)}. The answers for this dimension range from 1 - never heard of SBST before, to 4 - practical experience using SBST\@. Most participants had attended one lecture on SBST, that was part of the Verification and Validation course, and their answers reflect this. 
	\item \textbf{Statistics}. This dimension evaluates the participants' experience with statistics and the use of statistical methods. The answers range from no knowledge of statistics to practical experience using statistical methods to solve problems. Except for a few outliers, most participants had courses describing statistical methods, but little practical experience. 
	\item \textbf{Domain Familiarity}. This dimension evaluates the participants' familiarity with the domain, in this case with clustering in general and $k$-means clustering in particular. Most participants had no experience at all with clustering, with one participant having used clustering as a means of data analysis. 
\end{itemize}

The results show most participants have some expertise in software development, mostly at a theoretical level, due to the courses taken during their education. Most had no industrial experience in software development or testing. Most participants also believed they had a theoretical knowledge of the domain and some knowledge of statistics, again due to courses taken during their education. A basic introduction into SBSE was included in the course, so all participants had some basic knowledge of the technique.

We tried to isolate the method as the only variable that differed between the two groups of participants, to the extent to which this was possible. The relatively homogeneous level of knowledge and expertise lead us to conclude that any effects observed are due to differences between the ISBST and the manual tools, and therefore of the techniques, rather than differences in terms of experience between participants. 


\subsection{Test Cases} 
\label{sub:test_cases}

To answer \textbf{RQ1}, we looked at the test cases developed by the participants. During the course of the experiment, the 58 participants developed a total of $n_{total} = 4615$ test cases, of which $n_{auto} = 4154$ were developed and exported using the ISBST tool and $n_{manual} = 461$ were developed using the manual tool. This imbalance in terms of numbers is expected, as the ISBST tool generates a larger number of test cases in the same period of time than the manual method.

The analysis focuses on the behavior of the test cases. We defined the behavior of a test case as the set of scores obtained by that test case with respect to the objectives described in Table~\ref{tab:quality_objectives}.

A first look at the data consisted of performing the Mann-Whitney U test on each dimension of the test case behaviors. The Mann-Whitney is a non-parametric test, so no assumptions need to be made about the distribution of the population. The purpose of this analysis was to determine if there is a difference between test cases developed using the ISBST tool and the manually developed test with respect to each objective that describes the output. This initial analysis shows a statistically significant difference between the data developed by the two methods, with $p$-values of $< 10^{-5}$. The values for each objective can be found in Table~\ref{tab:initial_stats}. Thus, we can state that the test cases resulting from the two methods differ from each other with respect to all the objectives that define the output.

\begin{figure*}
	\centering
		\includegraphics[scale=0.55]{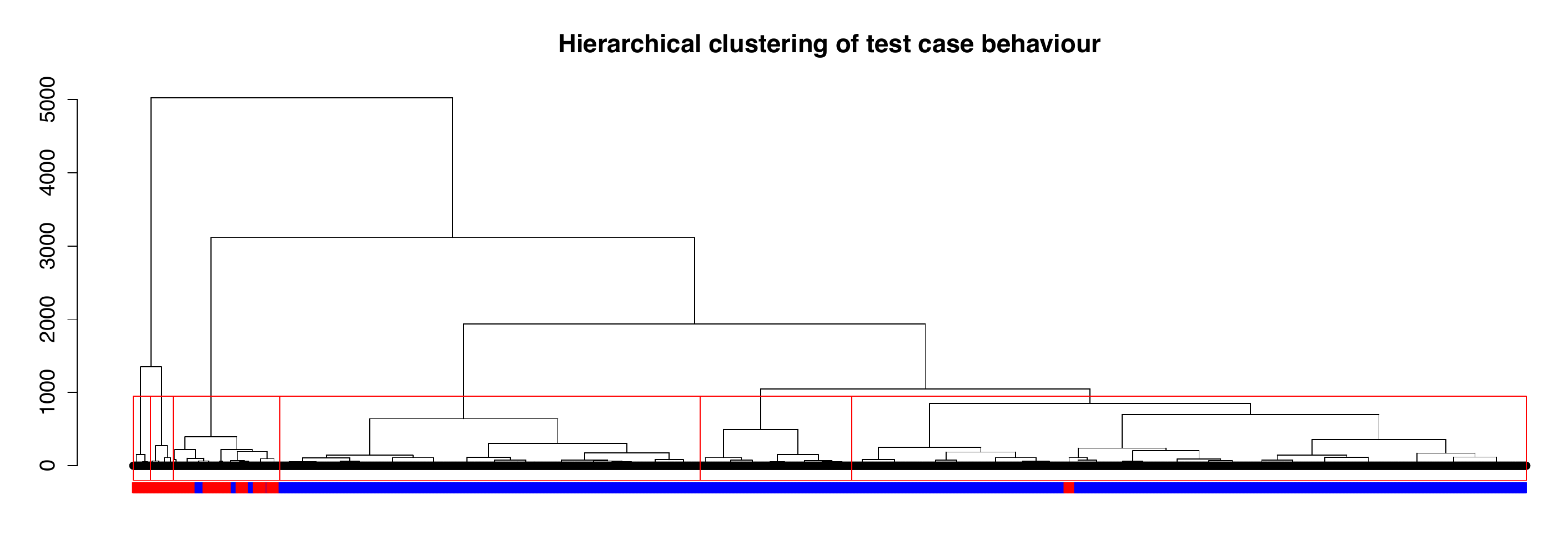}
	\caption{An overview of the hierarchical clustering. The y-axis is a measure of closeness of the clusters. Each leaf on the x-axis is a test case. They are arranged in order, based on the cluster assignment. The manually developed test cases are shown at the bottom of the image in red, the test cases developed using the ISBST system are in blue. The red boxes indicate the clusters used in the analysis. The width of each cluster is proportional to the number of test cases it contains. The left-most two clusters contain only manually developed test cases. The fourth and fifth cluster from the left only contain test cases developed by the ISBST system.}
	\label{fig:hier6clust}
\end{figure*}

To get a better understanding of how the test cases are distributed through the behavior space, we performed a hierarchical clustering on the objective scores. By clustering the test cases based on their behavior, we wanted to obtain distinct areas of the behavior space that were composed of similar test cases. We could then assess if there were any such areas that only had test cases resulting from one method. This would indicate that that region of the behavior space was explored by one method, but not the other. A cluster that contained test cases from both methods would indicate a region of the behavior space that both methods had explored. An overview of the results of the hierarchical clustering can be seen in Figure~\ref{fig:hier6clust}.

\begin{figure*}
	\centering
		\includegraphics[scale=0.45]{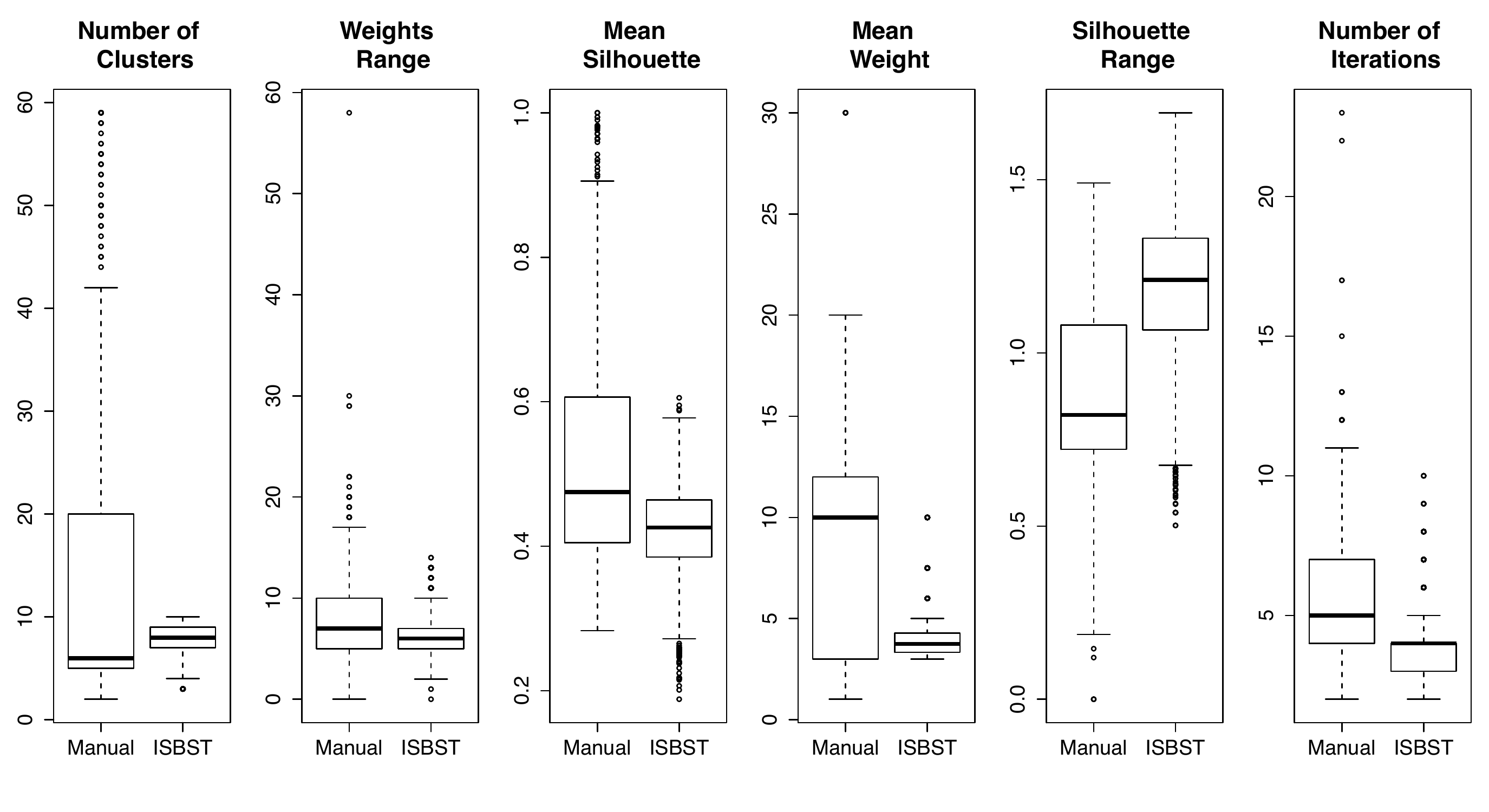}
	\caption{The difference between test case behavior (on each behavior dimension) between test case populations developed by the manual technique and the ISBST system.}
	\label{fig:rq1box}
\end{figure*}

\begin{table*}
	\center
	\scriptsize
	\begin{tabular}{ | p{0.09\linewidth} | p{0.09\linewidth}  | p{0.09\linewidth} | p{0.09\linewidth} | p{0.09\linewidth} | p{0.09\linewidth} | p{0.09\linewidth} | p{0.09\linewidth} | }	
		\hline
		  & Number of Clusters  & Number of Iterations & Mean Silhouette & Silhouette Range  & Mean Weight & Weights Range \\
		\hline
		Effect Size 	& 0.432	& 0.715					& 0.673 	& 0.217 		& 0.722 & 0.603 \\
		\hline
		Interpretation 	& negligible 	& medium		& medium	& large		& medium 	& small \\
		\hline
		$p$-value & \textless $10^{-5}$ & \textless $10^{-5}$  & \textless $10^{-5}$ & \textless $10^{-5}$ & \textless $10^{-5}$ & \textless $10^{-5}$ \\

		\hline	 
	\end{tabular}
	\caption{The $p$-values of the Mann-Whitney U test, for the differences (assessed in terms of behavior) between tests developed by the ISBST tool and those developed by the manual method. The effect sizes were calculated and interpreted using the Vagha-Delaney A measure.}	
	\label{tab:initial_stats}
\end{table*}

We cut off the hierarchical clustering after obtaining $n_{cluster} = 6$ clusters. The value was chosen arbitrarily, as it seemed to provide a finely grained view of the behavior space without increasing the complexity to unmanageable levels. 

The result, shown in Figure~\ref{fig:hier6clust}, shows the three larger clusters on the right of the image composed mostly of test cases resulting from the ISBST tool, with two of them being composed exclusively of test cases obtained by that method. The two clusters on the left of the figure are composed solely of test cases obtained by manual exploratory testing.

\begin{table}
	\center
	\scriptsize
	\begin{tabular}{ | p{0.12\linewidth} || p{0.3\linewidth} | p{0.3\linewidth} | }	
		\hline
		No. & Number of ISBST-generated test cases & Number of manually generated test cases \\
		\hline
		1 & 502 & 0 \\
		2 & 1392 & 0 \\
		\hline
		3 & 60 & 293 \\
		4 & 2200 & 35 \\
		\hline
		5 & 0 & 75 \\
		6 & 0 & 58 \\
		\hline	 
	\end{tabular}	
	\caption{The distribution of test cases to the clusters.}
	\label{tab:clus}
\end{table}

Table~\ref{tab:clus} shows the distribution of the number of test cases for each cluster. It can be seen that, while some overlap does exist, a large number of the test cases are in clusters that have no overlap. This seems to indicate that the two methods are focusing on different areas of the behavior space. 

\begin{figure}
	\centering
		\includegraphics[scale=0.5]{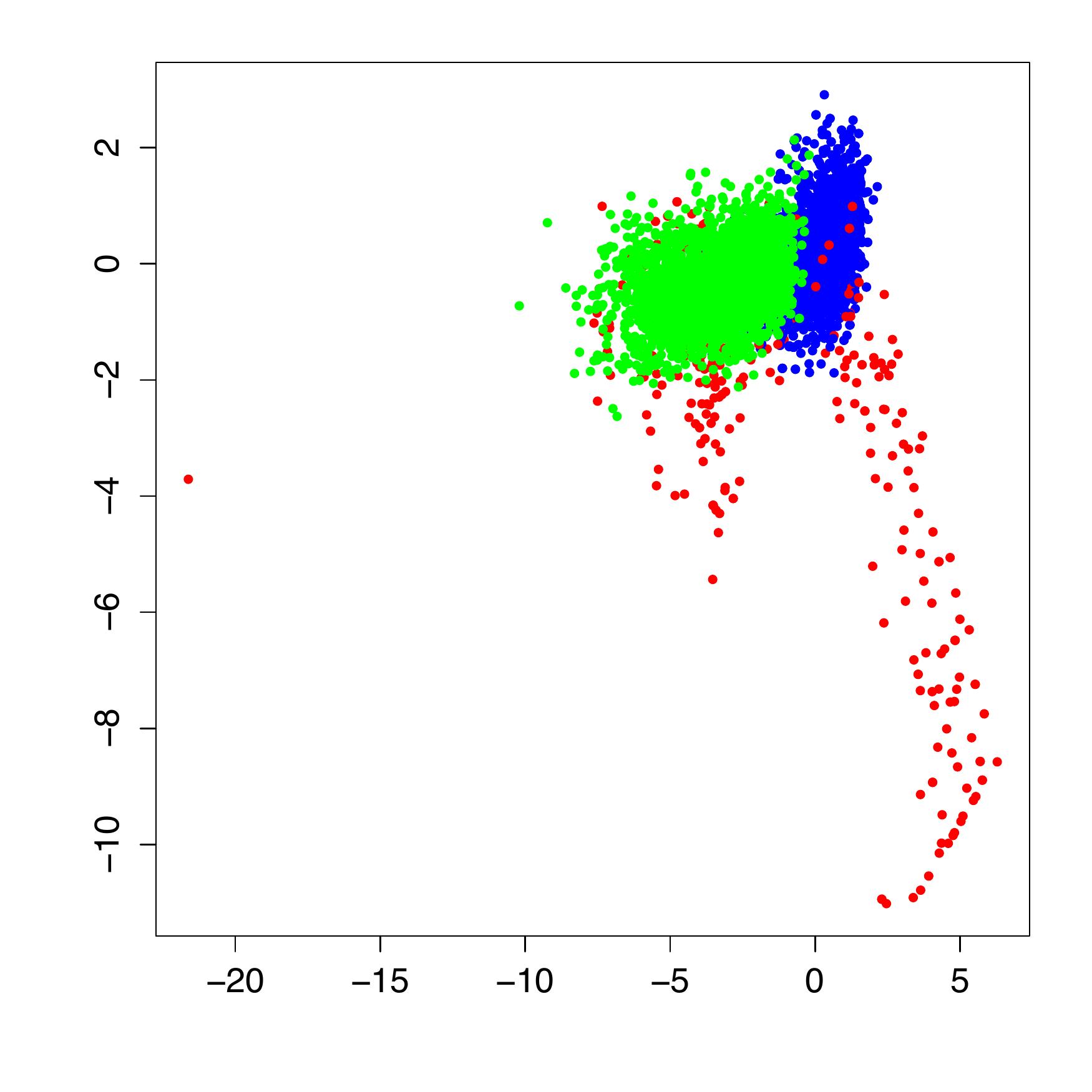}
	\caption{The results of a principal component analysis conducted on the test case behavior space. The tests obtained from the manual method are marked in red, those obtained from the ISBST tool are in blue, and the results of random generation in green. The upper right corner is where an ideal solution would have optimal values for all dimensions.}
	\label{fig:fig_pca2}
\end{figure}

These observations are also supported by the results of a principal component analysis, conducted on the behavior space. Three dimensions account for $64.88 \%$ of the observed variability. The most top two dimensions are shown in Figure~\ref{fig:fig_pca2}.

The individual test case behaviors were plotted and form distinct, though occasionally overlapping, clusters in the behavior space. The manually developed test cases (seen in red in Figure~\ref{fig:fig_pca2}) tend to be more spread out and cover a different area of the behavior space than those developed by the ISBST system (shown in blue in Figure~\ref{fig:fig_pca2}). The optimization objective for the search-based algorithm was the upper right corner of the graph. Both the manual technique and the ISBST system use the same objectives, i.e.\ the same behavior dimensions with the same directions for optimization. 

Thus, we can state confidently that the two methods investigate different areas of the behavior space. 

\begin{figure}
	\centering
	\includegraphics[scale=0.4]{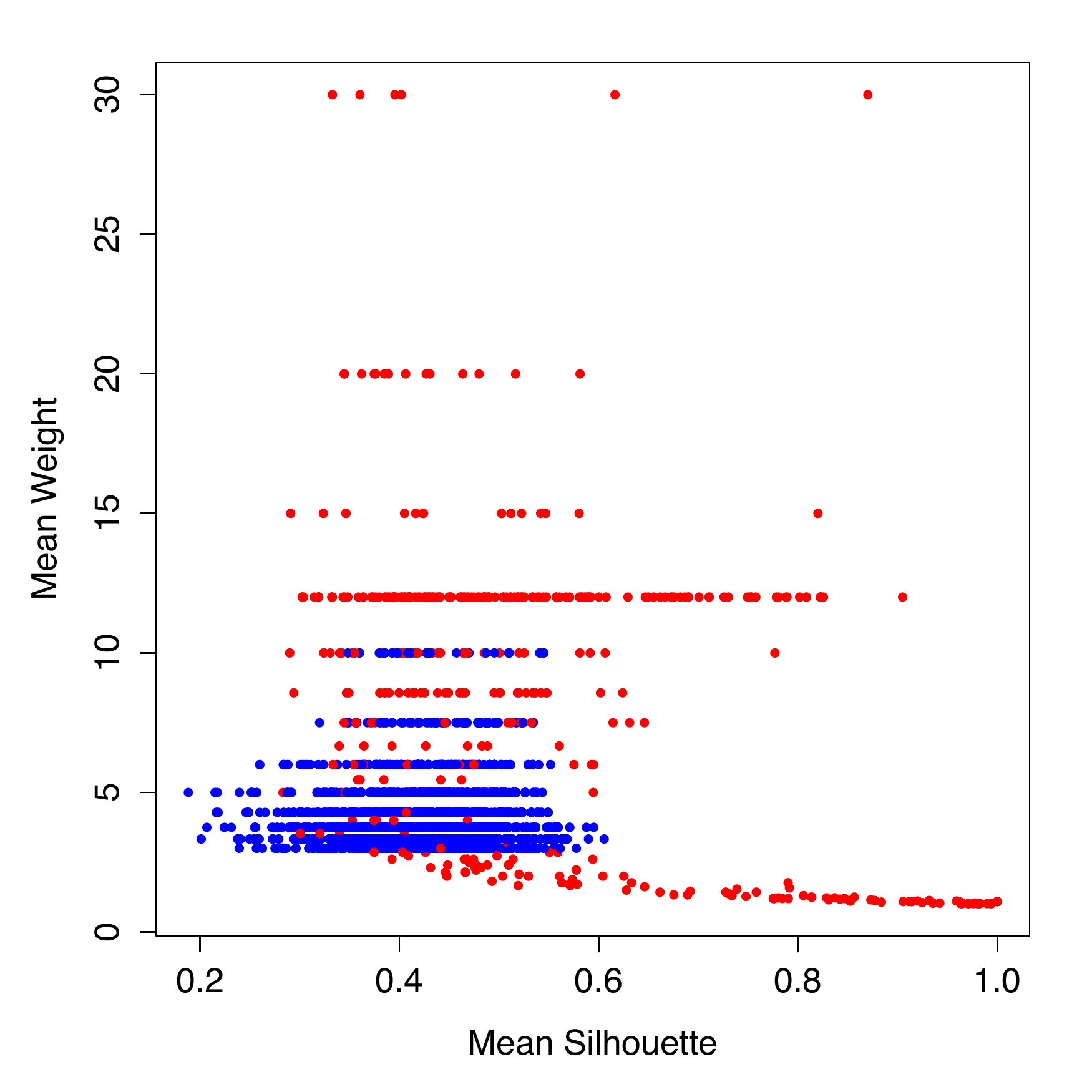}
	\caption{The behavior of the test cases with respect to Mean Silhouette and Mean Weight. The test cases obtained from the manual method are in red, those obtained from the ISBST tool are in blue. The origin of the graph (lower left corner)  is where an ideal solution would have optimal values for all dimensions.}
	\label{fig:fig_3objs_obvious1_ps}
\end{figure}

A closer analysis also shows differentiation in terms of the different objectives. Some of the objectives allowed the participants to find a front of solutions, illustrated in Figure~\ref{fig:fig_3objs_obvious1_ps} in red, even if that was dominated by the test cases developed by the ISBST system (shown in blue in the same figure). The dimensions in the figure were chosen because they best illustrate the solution front. 

\begin{figure}
	\centering
	\includegraphics[scale=0.4]{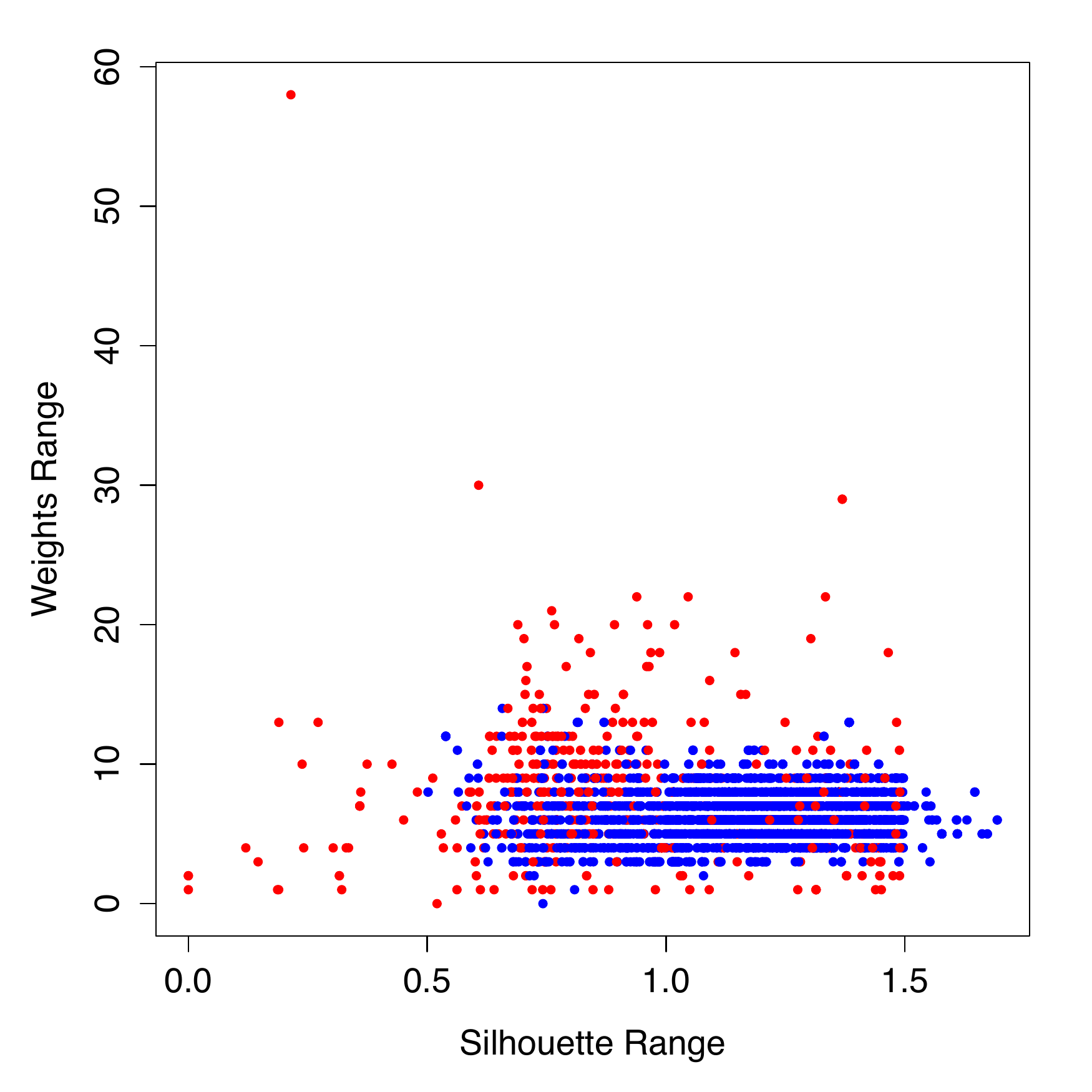}
	\caption{The behavior of the test cases with respect to Silhouette Range and Weights Range. The test cases obtained from the manual method are in red, those obtained from the ISBST tool are in blue. The top right corner of the graph is where an ideal solution would have optimal values for all dimensions.}
	\label{fig:fig_3objs_lessobvious_ps}
\end{figure}

For other objectives, e.g.\ Silhouette Range and Weights Range, the front is a lot less clear, as can be seen in Figure~\ref{fig:fig_3objs_lessobvious_ps}. In this case, the manually developed test cases are more random and show less evidence of a front. The two dimensions in the figure were chosen because they more clearly illustrate this.

A key element in the application of the ISBST tool is the complete and correct definition of objectives. In previous studies~\cite{marculescu2014initial} this was achieved by means of validating the objectives with domain specialists at the company. Such validation would, however, be impractical for a system aimed at a wider audience or when domain specialists are not available. Objectives that are more difficult to optimize, e.g.\ those in Figure~\ref{fig:fig_3objs_lessobvious_ps}, may hide improvements in other selected objectives. 

As a result of this analysis, we can conclude that the answer to the first research question is that the two methods, the ISBST tool and the manual exploratory testing, investigate different areas of the behavior space. We have found that there are regions of the behavior space that are only explored by only one of the two methods.


\subsection{The Effect of Interaction} 
\label{sub:the_effect_of_interaction}

The ISBST tool is composed of two elements: the search component (identified in Section~\ref{sub:interactive_search_based_software_testing_isbst_} as the Inner Cycle) and the interaction component (called Outer Cycle). The results up to this point have shown that the ISBST tool, taken as a whole, achieves the goal of investigating areas of the search space that the manual method does not reach.

The effect of the search component of the ISBST tool is the difference between the initial population at the beginning of the search, and the population at the end of the non-interactive execution.

\begin{figure*}
	\centering
		\includegraphics[scale=0.45]{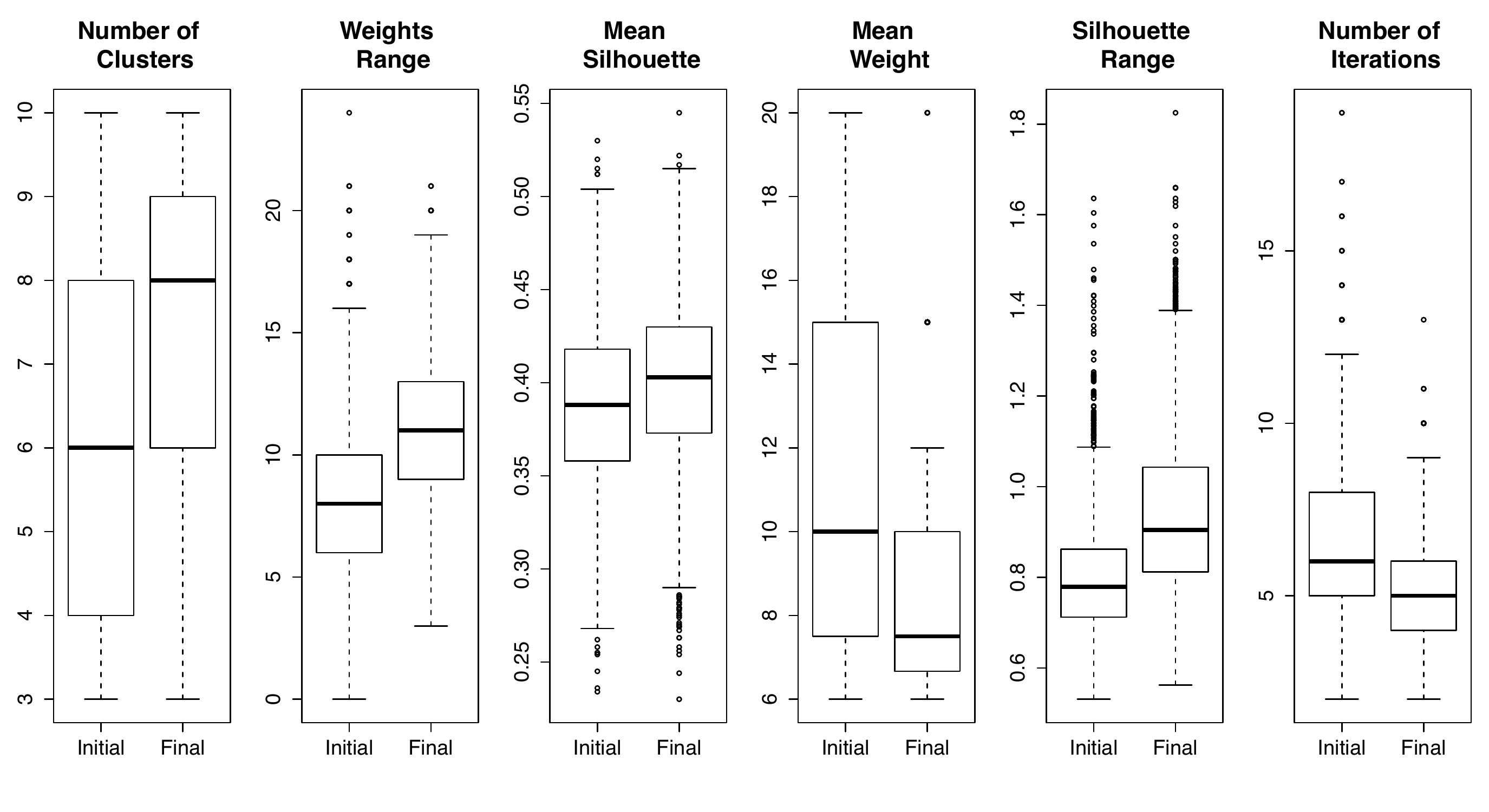}
	\caption{The effect of search on the test case population. For each search objective, the values for the initial population are marked with ``Initial'', and those from the final population with ``Final''. Note: each search objective is shown to a different scale.}
	\label{fig:images_search-effect}
\end{figure*}

\begin{figure*}
	\centering
		\includegraphics[scale=0.45]{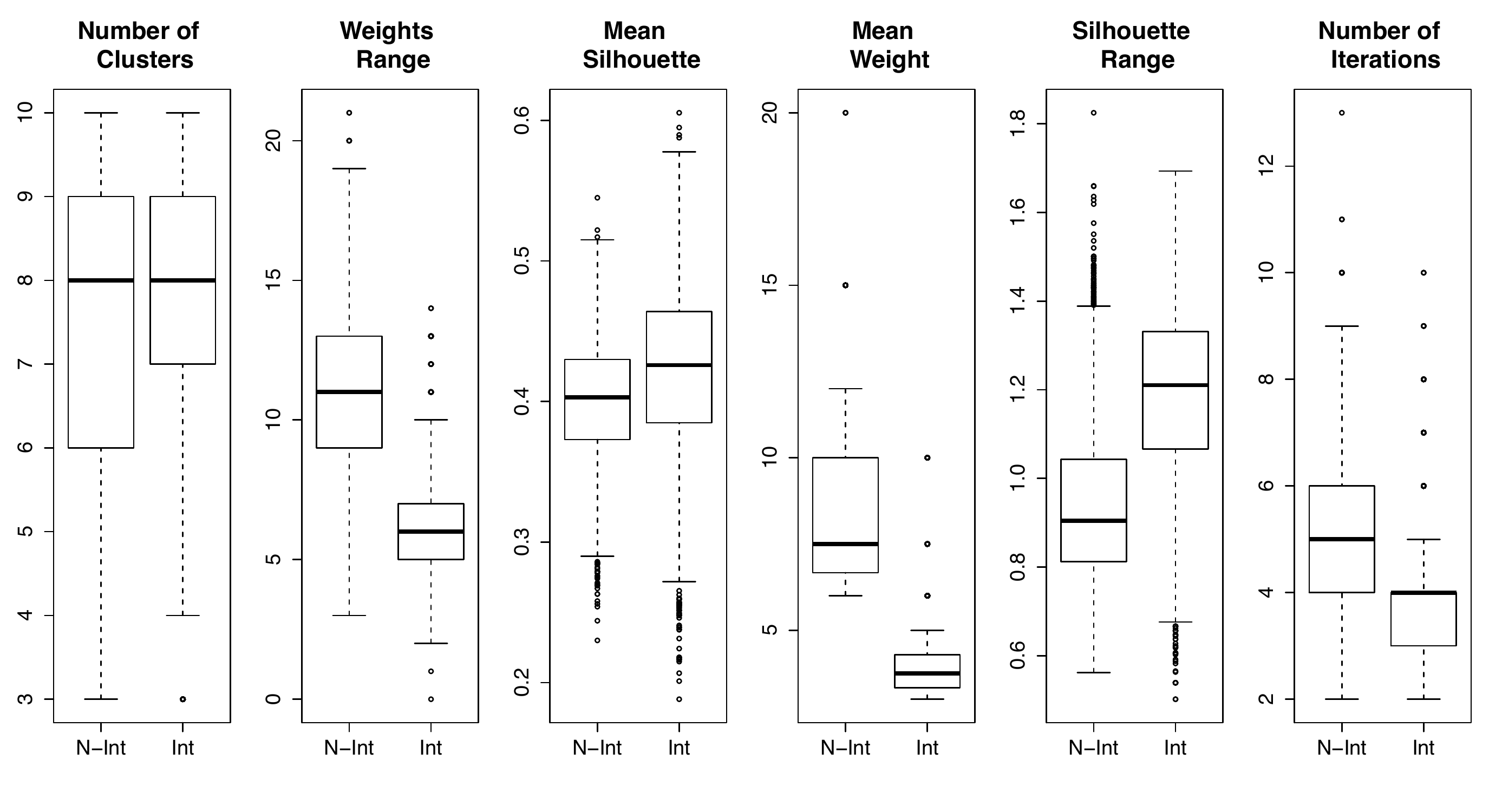}
	\caption{The effect of interaction on the test case population. For each search objective, the values for the non-interactive execution are marked with ``N-int'', and the values for execution including interaction are marked ``Int''. Note: The scale for each search objective is different. All values are for the final population at the end of both the interactive and non-interactive runs.}
	\label{fig:interaction-effect}
\end{figure*}

Figure~\ref{fig:images_search-effect} shows the difference between the initial and final populations in the non-interactive execution. It is clear that for those objectives that were to be maximized (Weights Range, Mean Silhouette, and Silhouette range) the final population shows higher values than the initial one. Conversely, for the objectives that were to be minimized (Mean Weight and Number of Iterations) the final population scores are lower. The one exception is the Number of Clusters, that should be minimized, but overall shows higher values. This is due to the fact that less optimal values are needed for this behavior attribute in order to improve the others. To test the overall statistical significance of the change in values, the Mann-Whitney U test was conducted on the test case behaviors. The null hypothesis, that the samples come from the same population, was rejected for all the search objectives with $p < 10^{-5}$. The values can be seen in Table~\ref{tab:interaction_effect}. We can confidently conclude that the search component has a considerable effect on the outcome of the ISBST tool.

\begin{table*}
	\scriptsize
	\begin{tabular}{ | p{0.2\linewidth} || p{0.08\linewidth}  | p{0.08\linewidth} | p{0.08\linewidth} | p{0.08\linewidth} | p{0.08\linewidth} | p{0.08\linewidth} | p{0.08\linewidth} | }	
		\hline
		Test & Number of Clusters  & Number of Iterations & Mean Silhouette & Silhouette Range  & Mean Weight & Weights Range \\
		\hline

 		Search Effect Sizes &	0.355 & 0.659 & 0.408 & 0.242 & 0.644 & 0.237 \\
		\hline
		Search Effect Interpretation & small & small & small & large & small & large \\
 		\hline
		Search Effect Significance & \textless $10^{-5}$  & \textless $10^{-5}$ & \textless $10^{-5}$ & \textless $10^{-5}$ & \textless $10^{-5}$ & \textless $10^{-5}$ \\
		\hline
		
		Interaction Effect (actual) Sizes & 0.454 & 0.779 & 0.377 & 0.185 & 0.978 & 0.918 \\
		\hline
		Interaction Effect (actual) Interpretation & negligible & large & small & large & large & large \\
		\hline 
		Interaction Effect Significance (actual)  & \textless $10^{-5}$  & \textless $10^{-5}$ & 0.5098 & \textless $10^{-5}$ & \textless $10^{-5}$ &  0.000301 \\
		\hline
		
		Interaction Effect (potential) Sizes & 0.423 & 0.923 & 0.392 & 0.983 & 0.184 & 0.754 \\
		\hline
		Interaction Effect (potential) Interpretation & small & large & small & large & large & large \\
		\hline		
		Interaction Effect Significance (potential) & \textless $10^{-5}$  &  \textless $10^{-5}$ & 0.03252 & \textless $10^{-5}$ & 0.6022 & \textless $10^{-5}$ \\
		\hline	 
	\end{tabular}
	\caption{The $p$-values of the Mann-Whitney U test, for the differences between the overall strategy of the participants and the non-interactive execution. The Search Effect Significance shows the significance of differences between the initial (random) test case population and the final test case population (after a non-interactive ISBST run). The Interaction Effect values show significance of the differences between the final population of test cases for the interactive and non-interactive runs of the ISBST. Effect sizes were calculated and interpreted using the Vargha-Delaney A measure.}	
	\label{tab:interaction_effect}
\end{table*}

To assess the impact of the interaction, we compared the populations of test cases resulting from the non-interactive execution against the test data resulting from the experiment. Two comparisons were conducted on an objective by objective basis, and the values for the two populations can be seen in Figure~\ref{fig:interaction-effect}. Note that interaction seems to be further from the optimum than running with a fixed set of objectives. A non-interactive run supposes that all the objectives, and their relative importance, are known in advance and fixed, which is difficult to achieve \emph{a priori}.  

We distinguish between Actual Interaction Significance and Potential Interaction Significance, both seen in Table~\ref{tab:interaction_effect} as Interaction Significance (actual) and Interaction Significance (potential), respectively. For the Actual Interaction Significance, we compared the test cases exported by the participants in the experiment against those developed by the ISBST system without any interaction. This has the benefit of comparing the test cases that the participants though best against the non-interactive run, but gives little information about how the two test case populations compare. 

To compare the entire population produced by the participants using the ISBST system against that developed by the ISBST system without the benefit of interaction, the Potential Interaction Significance was calculated, and is seen as Interaction Significance (potential) in Table~\ref{tab:interaction_effect}. This provides more information on the overall difference between the two populations of test cases. 

In spite of the overall variance in the interaction strategies and approaches used by the participants, the interaction component clearly has a significant impact on the overall outcome. 

A closer inspection of data available for individual participants provides more evidence to support the notion that some of the objectives are more intuitive and easier to optimize. It also emphasizes the importance of search objective definition in the application of the ISBST tool. 


\subsection{Fatigue} 
\label{sub:participants_and_fatigue}

To answer \textbf{RQ3}, we asked each participant to fill in the NASA Task Load Index after completing each of the two 45-minute sessions. The Mann-Whitney U Test was performed on the results to determine if any of the differences observed were statistically significant. 

\begin{figure*}
\centering
	\subfigure{\includegraphics[scale=0.45]{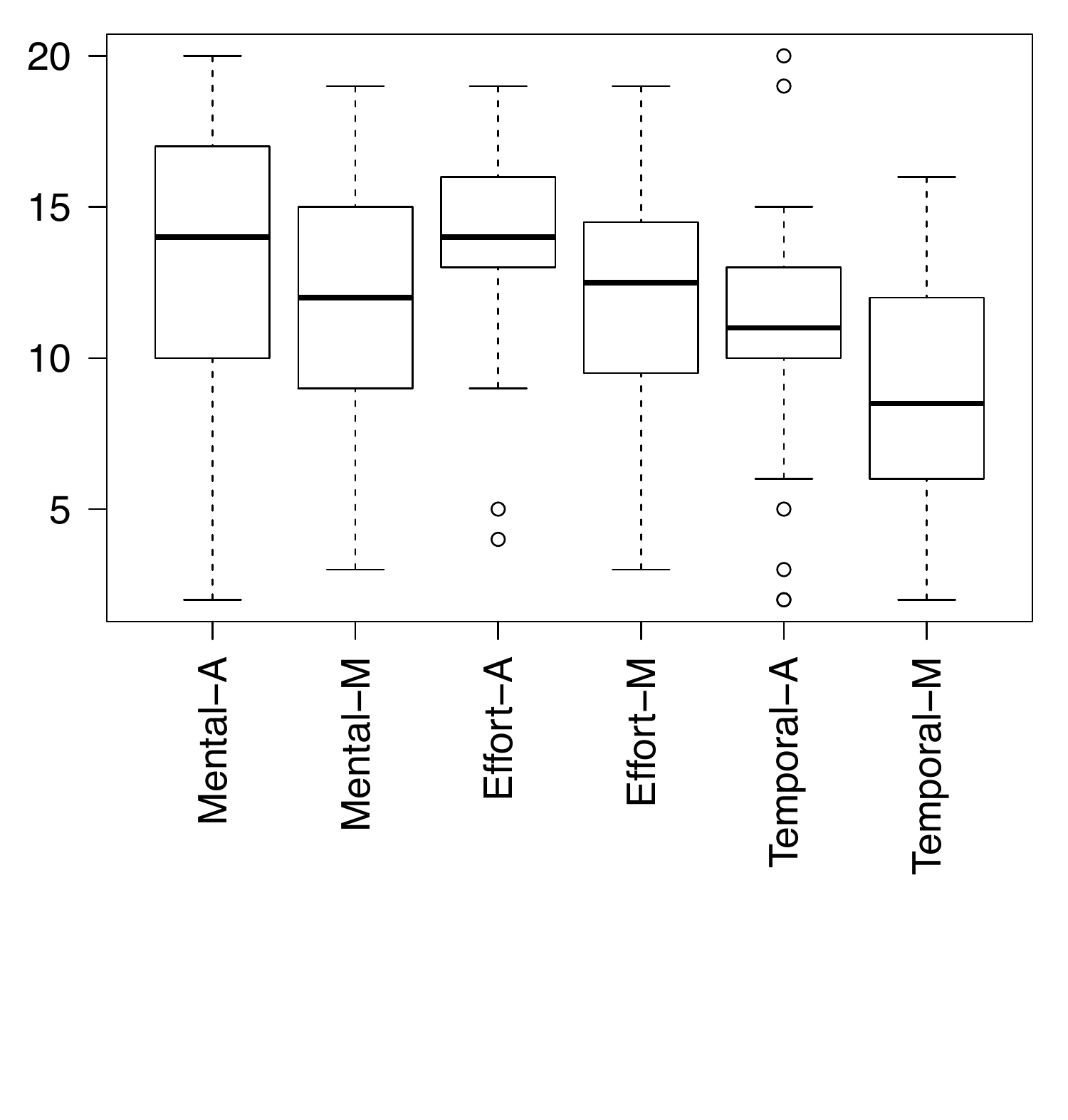}}
	\subfigure{\includegraphics[scale=0.45]{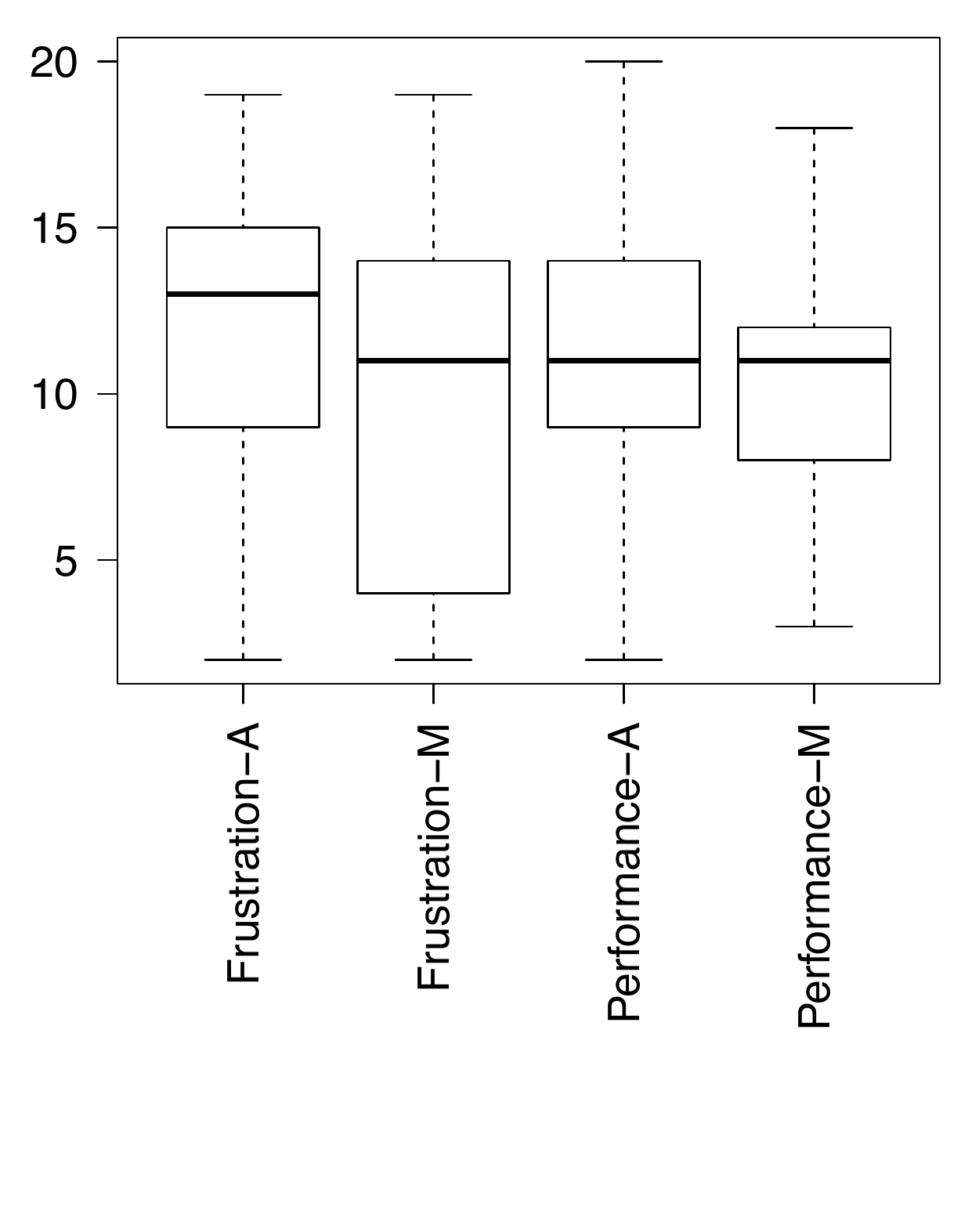}}
	\caption{The level of demand the participants felt was placed upon them by the two methods and the perceived level of performance, as self-assessed using the NASA-TLX after the second 45-minute session. All measurements are on the same scale. The results for the ISBST tool are marked with A - automated, and those for the manual technique are marked with M - manual.}
	\label{fig:fatigue}
\end{figure*}

\begin{table*}
	\scriptsize
	\center
	\begin{tabular}{ | p{0.13\linewidth} || p{0.12\linewidth} |  p{0.12\linewidth} |  p{0.12\linewidth} |  p{0.12\linewidth} |  p{0.12\linewidth} | }	
		\hline
		Session & Mental Demand & Effort & Temporal Demand & Frustration & Performance \\
		\hline
		
		After the first 45-minute session  & \textbf{2.6} & \textbf{3.12} & 2.4 & 0.38 & 0.13 \\
		\hline
		After the first 45-minute session  & 1.03 & \textbf{2.33} & 1.93 & 2.31 & 0.9 \\
		\hline	 
	\end{tabular}
	\caption{The differences between the means of the NASA-TLX values for the manual and ISBST tool. The positive values show that the ISBST tool is more demanding than the manual technique, with the statistically significant differences are marked in bold.}	
	\label{tab:fatigue3}
\end{table*}

The results show that, overall, the participants perceived the ISBST tool to be more demanding on their mental, effort, and temporal resources. It is interesting to note, however, that in terms of the frustration engendered by the each of the methods, and of the performance (as perceived by the participants themselves), the differences were not statistically significant. 

After the first session, the ISBST system was assessed as being more mentally demanding ($p = 0.013 $) and more demanding in terms of Effort ($ p = 0.009 $). The data collected at the end of the second 45-minute session shows an interesting effect: only the difference in terms of Effort is still statistically significant ($ p = 0.022 $). 

At the end of the second session, each participant had had the experience of using both methods, and was, thus, in a better position to compare them. A more detailed view of the results can be found in Figure~\ref{fig:fatigue}, and difference between the means is shown in Table~\ref{tab:fatigue3}. 

The statistically significant differences are marked in bold.

As a result, we can conclude that the answer to \textbf{RQ3} is that the ISBST tool is, indeed, more demanding of the domain specialist, at least initially. It is worth noting, however, that the increased effort did not result in significantly higher frustration, and resulted in similar performance.


\FloatBarrier


\section{Discussion} 
\label{sec:discussion}

Our findings suggest that the ISBST system and the manual exploratory testing technique investigate different regions of the SUTs behavior space. Therefore, we conclude that the two methods are complementary and that the diversity of the test suite can be increased by combining relevant test cases produced by the two methods.

We assess the two methods in terms of the regions of the behavior space that they cover, and therefore, in terms of diversity. We cannot state, in general, that this is a good measure for the number of faults found by the test suite. However, there is evidence to suggest that increase diversity does result in more fault-revealing test suites~\cite{cognitively_diverse, diameter}.  

A closer look at the data shows that the test cases developed by the ISBST tool seem somewhat more concentrated, while those developed by the manual technique seem more diverse. Thus, it would seem that adding the test cases produced by the ISBST system would only result in a small increase in the diversity within the test case population. We will argue that the test cases developed by the ISBST tool are a useful addition to the test suite. 

The first issue that needs to be discussed is that of relevance. The requirements we defined to develop and analyze the test cases, as well as each of the direction of each search objective, were arbitrary. This makes it difficult to assess how relevant the test cases are. But it is worth noting that many of the manually developed solutions that are far from the automated set, are also test cases that the ISBST system would evaluate as having low fitness. 

As mentioned in the previous sections, the search-based system has a set of search objectives, each objective with a direction, that is uses to optimize. Participants could choose which of those objectives to favor, but all selected objectives would be optimized according to those directions. This is a limitation that the manual technique did not have, allowing participants greater freedom to explore test cases that the ISBST system would have dismissed as having too low a fitness value. 

The evidence suggests that, when faced with a clear set of objectives, the ISBST system is better at evolving solutions that get close to those objectives. As discussed in Section~\ref{sec:results}, there are dimensions where a clear front of non-dominated solutions is obtained, and that front is closer to the optimum for solutions developed by the ISBST tool. Such a method, however, is more vulnerable to incomplete and inaccurate objectives. 

This supports the notion that the two methods are complementary, rather than competing. Focusing on the objectives does, however, limit the ISBST tool. If validating the search objectives is not possible or not practical, alternative mechanisms should be found, that can conduct a search by focusing on exploration rather than optimization. There is a potential to improve the ISBST tool by incorporating mechanisms to increase diversity, e.g.\ Novelty Search~\cite{Lehman:2011:EDV:2001576.2001606}, Viability Evolution~\cite{EPFL-ARTICLE-191284}, or MAP-Elites~\cite{MouretC15}.

In analyzing the data regarding \textbf{RQ2}, we note that some behavior attributes show greater variation or greater improvement. This can be due to several causes. 

First, the difficulty in optimizing each attribute varies, as does their scale. Some attributes have larger scales, where changes in values are more easily observed. Some objectives are easier to optimize and, therefore, can reach more obvious improvements with less effort. Moreover, objectives are sometimes contradictory. The Number of Cluster objective is a good example, as worse fitness values are needed for that quality objective in order to obtain better fitness overall. 

In addition, the interactive component highlights objectives that show the most improvement. 
This can be compounded by the interactive component, as objectives that show the most improvement will also draw more attention from the domain specialists and receive more of their time and focus. The data we presented and discussed in specific to the SUT used in this study, but we expect that in any domain, such differences will arise. 

Note also that, as the data analyzing the difference between interactive and non-interactive search, in Section~\ref{sub:the_effect_of_interaction}, shows that the interactive search seems to be worse overall than the non-interactive search. This is to be expected, as knowing all the objectives ahead of time, and having that selection and weighting stable throughout the search leads to better optimization. As mentioned before, however, coming up with a stable set of objectives, properly weighted is difficult to do \emph{a priori}. Moreover, as the search progresses, the relative importance of objectives may shift, or new search objectives may be added. Thus, the reason for introducing interaction is giving the ISBST tool the flexibility to adapt to changes in objectives or in their relative importance.

Our discussion of fatigue, and our answer to \textbf{RQ3}, showed that participant did consider the ISBST tool to be more demanding mentally and in terms of effort and time requirements. First, this is to be expected, as the added complexity of the search increases the distance between the participants' actions and their effect on the SUT's behavior and the resulting test cases. 

In spite of the added fatigue, however, we would like to note that all participants were able to use both techniques to create test cases. Therefore, we surmise that the increased demands placed on the participants by the ISBST tool did not prevent the participants from completing their tasks.


\section{Threats to Validity} 
\label{sec:threats_to_validity}

There are several threats to the validity of this study that need to be discussed. 

\subsection{Construct Validity} 
\label{sub:construct_validity}

First, we propose the ISBST system as a complementary technique, to quickly explore the behavior space, based on a set of desired behavior attributes. Thus, our study focused on diversity of test cases and on assessing how the techniques we used explored different regions of the behavior space. This relies on the assumption that a more diverse set of test cases will also result in better fault finding.  

We cannot claim that using the ISBST tool as a complement to existing testing techniques will result in greater fault finding or increased quality. Only that such a tool increases the number of SUT behaviors that are being investigated. There is, however, work to suggest that test case diversity does result in greater structural and fault coverage~\cite{diameter}. Further studies are necessary before any conclusions can be drawn regarding the impact of such methods on software quality in general and on fault finding in particular. 


\subsection{External Validity} 
\label{sub:external_validity}

A second issue is the choice of SUT\@. For this study, we wanted a SUT with a high-dimensional input and behavior, and where assessing the quality of the solution would involve human participants. The chosen SUT had to strike a balance: it had to be simple enough to understand and use in the limited time available to the participants. However, it also had to be complex enough to benefit from exploratory black-box techniques and not allow optima to be simply calculated or exhaustively searched. We chose a system with a high-dimensional input and behavior, based on our previous experience with our industrial partner. The $k$-means clustering algorithm has the complexity of some of the embedded modules our industrial partner works with, without being domain specific. 

The results of the algorithm can be described as the behavior used to optimize, but participants also had access to the assignment of points to clusters. This allowed them to interpret a test case candidate both from the objective perspective of the computed behaviors and from a subjective evaluation of whether or not the clustering seemed valid to them. Thus, evaluating candidate test cases could only be done by involving human participants. 

Nevertheless, we cannot claim our conclusions can be applied to any SUT\@. Systems that have different characteristics: with a lower number of inputs or where the behaviors are not dependent on the inputs alone, could show different effects. Moreover, systems that do not require human input to assess candidate solutions might not need the interactive component altogether. Before using this technique in new domains or where such characteristics of the SUTs are not known, further validation is required. 

Finally, the choice of participants and their level of expertise are also threats to validity. The participants for the experiment were all students at one university in Sweden, and participating in the Verification and Validation course, part of the Software Engineering master's program. This may result in the participants not being representative for the domain specialists they stand in for. The participants are software engineers in training. While they do not have the experience, their skills and knowledge are relevant to the experiment. In addition, work by Kuzniarz et al.\@~\cite{kuzniarz2003students} suggests that conclusions can be drawn if students are less familiar with a new method being proposed than to the standard it is compared against. Moreover, H{\"o}st et al.\@ suggest that conclusions drawn from experiments with students can hold if carried out with students in their final year~\cite{usingStudents}. Both these assumptions hold in our case. Nevertheless, further validation is required, to assess the impact of experience on the ISBST system.

An added concern is that of domain knowledge. The ISBST tool relies on domain specialists to assess candidate solutions and guide the search, based on their knowledge and expertise. The participants in our experiment were chosen based on their willingness to participate and their availability in the numbers required for the experiment, rather than domain knowledge or experience. The initial survey of the participants also confirms that their knowledge of the problem domain, self-evaluated, is limited. This suggests that experienced domain specialists may exhibit different behavior and may obtain different results. 

Thus, before applying the findings in industry, further validation is required. Prototyping any tools with the domain specialists would provide improvements to the tool itself, as well as greater insight into any domain or context specific limitations that could influence the search.



\section{Conclusions} 
\label{sec:conclusions}

In this paper we have presented an experiment comparing an implementation of interactive search-based software testing, the ISBST tool, and a manual exploratory black-box technique, in terms of developing test cases for a given SUT\@. The SUT in this case was the Julia implementation of a $k$-means clustering algorithm. 

\textbf{RQ1} was concerned with the degree to which test cases developed by the automated method investigate different regions of the behavior space, thereby increasing the diversity of available candidates.

The experiment has shown that the two methods focus on different areas of the behavior space and enabled participants to create test cases that exercise different types of behaviors of the system. This indicates that the ISBST tool is a useful complement to the exploratory black-box technique: it allows participants to develop test cases that investigate different behaviors and characteristics of the system. 

\textbf{RQ2} focused on identifying if both the interactive and the search components of the ISBST system contribute to the observed effects.

To answer this question we have conducted a laboratory experiment to evaluate the impact of both the interaction and the search component on the outcome of the ISBST system. We conclude that both components significantly influence the search process and the final outcome. 

The subject of \textbf{RQ3} was the demand placed on the domain specialists by each of the two methods. We conclude that the ISBST system seems to demand more effort from the domain specialist, at least initially, but does not result in a significant increase in frustration or a significant degradation of performance.

An additional conclusion is that there are limitations to the objective-based approach that the ISBST tool uses to guide the search. The objectives used, if incomplete or improperly formulated, can be biased against certain types of behaviors and thus limit the search. This is not a problem if the objectives can be extensively validated with domain experts and constantly updated. If such validation is not possible or not affordable, an alternative method, e.g.\@ non-competitive or exploration focused evolutionary computation, could provide a way to mitigate the limitations of ISBST\@. Future work will focus on assessing exploration focused algorithms and investigating the benefits they may provide


\bibliographystyle{elsarticle-num}
\bibliography{p2}

\end{document}